\documentclass[aps,prx,showpacs,superscriptaddress,twocolumn,floats,epsfig,longbibliography]{revtex4-2}
\usepackage{amssymb}
\usepackage{amsbsy}
\usepackage[final]{microtype}
\usepackage{amsmath}
\usepackage{epsfig}
\usepackage{color}
\usepackage{float}
\usepackage{xr-hyper}
\usepackage[colorlinks,linktocpage,bookmarks=false,citecolor=blue,linkcolor=red,urlcolor=blue]{hyperref}
\usepackage{tikz}
\usepackage{quantikz}

\newcommand{\beg}{\begin{gather}}
\newcommand{\eeg}{\end{gather}}
\newcommand{\beq}{\begin{equation}}
\newcommand{\eeq}{\end{equation}}
\newcommand{\bea}{\begin{eqnarray}}
\newcommand{\eea}{\end{eqnarray}}

\newcommand{\be}{\begin{equation}}
\newcommand{\ee}{\end{equation}}
\def\ba{\begin{aligned}}
\def\ea{\end{aligned}}
\def\bes{\begin{subequations}}
\def\ees{\end{subequations}}
\def\bal{\begin{align}}
\def\eal{\end{align}}

\newcommand{\rg}[1]{{\color{black}#1}}
\newcommand{\rgg}[1]{{\color{black}#1}}

\usepackage{soul}
\usepackage{subfigure}
\usepackage{ulem} 
\usepackage{hyperref}
\graphicspath{{./figures/}}

\begin{document}
\title{\rgg{Enhancement} of quantum annealing via n-local catalysts}

\author{Roopayan Ghosh}
\thanks{roopayan@iitbbs.ac.in}
\affiliation{Department of Physics and Astronomy, University College London, Gower Street, WC1E 6BT, London, United Kingdom}
\affiliation{School of Basic Sciences, Indian Institute of Technology, Bhubaneswar, 752050, India }
\author{Luca A. Nutricati}
\affiliation{London Centre for Nanotechnology,  University College London, Gower Street,  WC1E 6BT, London, United Kingdom}
\author{Natasha Feinstein}
\affiliation{London Centre for Nanotechnology,  University College London, Gower Street,  WC1E 6BT, London, United Kingdom}
\author{P. A. Warburton}
\affiliation{London Centre for Nanotechnology,  University College London, Gower Street,  WC1E 6BT, London, United Kingdom}
\author{Sougato Bose}
\affiliation{Department of Physics and Astronomy, University College London, Gower Street, WC1E 6BT, London, United Kingdom}
\begin{abstract}
The \rg{potential} quantum speedup in solving optimization problems via adiabatic quantum annealing is often hindered by the closing of the energy gap during the anneal, especially when this gap scales exponentially with system size. In this work, we alleviate this bottleneck by demonstrating that for the NP-complete Maximum Weighted Independent Set (MWIS) problem, an informed choice of $n-$local catalysts (operators involving $n$ qubits) can re-open the gap during the \rg{annealing} process.
By analyzing first-order phase transitions in \rg{toy} instances of the MWIS problem, we first identify direct-tunneling catalysts that effectively eliminate the transition and provide an analytical discussion on when the sign of the catalyst influences its impact. We then reveal that $n-$local catalysts exponentially improve gap scaling and in certain scenarios are as effective as direct tunnel coupling between two minima. \rg{Utilizing this understanding, we show that they also increase the efficiency of ground state preparation via adiabatic quantum annealing in random graphs and analytically demonstrate the necessity of their placement across unfrustrated loops in the graph for effective performance in MWIS problems}. Additionally, using a circuit implementation of the $n$-local catalyst (requiring $2n$ nearest-neighbour gates), we demonstrate that both the circuit depth and the total number of gates required to solve the problem are reduced \rg{by several orders of magnitude when compared to the discrete-time version of traditional quantum annealing using local drivers.}  Our analysis suggests that non-local quantum fluctuations entangling multiple qubits as a catalyst are key to achieving the desired quantum advantage.
\end{abstract}
\maketitle
\section{Introduction}

One of the primary goals of quantum computation is to solve combinatorial optimization problems more efficiently than classical computers. This can be achieved by mapping the problem to the ground state solution of a suitable Hamiltonian. The complexity of the problem is then transferred to the simulation of the ground state of this Hamiltonian on a quantum simulator. One method to simulate such a state is via adiabatic quantum annealing, which is a method where one starts from an easily preparable ground state of a known Hamiltonian and adiabatically varies the potential landscape to reach the target Hamiltonian \cite{10.1063/1.2995837,Johnson2011,RevModPhys.90.015002,RevModPhys.80.1061,Jiang2018}. According to the adiabatic theorem, this ensures that the correct state, and thus the solution, is obtained if the procedure is applied slowly enough. However, in practice, the chosen pathway can encounter points where the system undergoes a phase transition, leading to regions with extremely small energy gaps between the ground and first excited states~\cite{dutta2010quantum}. This threatens adiabaticity, as excitations can easily be generated in these regions. Energy gaps which are exponentially small in system size, a common feature for first-order phase transitions and in some other scenarios~\cite{BORGS1993168,PhysRevLett.109.030502,PhysRevResearch.5.043236}, are particularly problematic, driving the time complexity to solve the problem to be exponential in system size, thus negating any potential quantum advantage.

Recent research has therefore focused on strategies to avoid such scenarios. Promising approaches include counter-diabatic driving\cite{PhysRevLett.111.100502, PhysRevX.4.021013, doi:10.1073/pnas.1619826114, Iram2021,
PRXQuantum.4.010312,  schindler2024counterdiabaticdrivingperiodicallydriven}, the quantum approximate optimization algorithm (QAOA)\cite{farhi2014quantumapproximateoptimizationalgorithm,Farhi2022quantumapproximate,BLEKOS20241,PRXQuantum.2.010309}, quantum walks (QW)\cite{doi:10.1080/00107151031000110776,Venegas-Andraca2012,doi:10.1137/090745854,Wadhia2024,gerblich2024advantagesmultistagequantumwalks,Imparato_2024}, \rg{optimal control techniques~\cite{Pecci_2024}}, inhomogeneous driving\cite{doi:10.7566/JPSJ.87.023002,PhysRevA.98.042326,Adame_2020}, restricting evolution to certain symmetry sectors\cite{Bernaschi2024}, the addition of non-stoquastic interactions\cite{Seki2012,Seoane_2012_2, crosson2014differentstrategiesoptimizationusing, Seki_2015,Nishimori2017_2, PhysRevB.95.184416,PhysRevA.99.042334,PhysRevResearch.3.043013}, and the use of directed catalysts\cite{PhysRevA.103.022608,feinstein2023}. While each of these methods typically reduces computational complexity for certain classes of problems, a universal protocol that works across all types of problems is yet to be developed.

In this work, we show a universal way to eliminate a class of first-order phase transitions during quantum annealing while preserving adiabaticity: multi-qubit fluctuations induced by $n-$local terms coupling $n$ qubits simultaneously.  Previous research has suggested the usage of such terms in drivers,~\cite{Matsuda_2009,Mazzola_2017}, \rg{i.e. quantum fluctuations which are gradually reduced during the course of the anneal, but they have limited impact in certain scenarios,} such as in uniformly sampling degenerate ground states~\cite{PhysRevA.100.030303}. We propose \rg{their} implementation as catalysts—additional quantum fluctuations introduced during the anneal but absent at both the start and end. This approach introduces a distinct energy scale compared to the driver, enabling traversal of alternative annealing pathways and easier initial state preparation regardless of the \rgg{sign of the additional couplings}. Additionally, our concern is speeding up the anneal process by removing the phase transitions in problems typically with a non-degenerate solution.~\footnote{For problems with exactly degenerate classical solutions, the minimum gap clearly occurs only at the end of the anneal; while our method can still mitigate perturbative crossings during the evolution, it cannot guarantee equal weight across all degenerate ground states, and hence we consider a non-degenerate solution.} Consequently, we identify \textit {what} constitutes an effective catalyst, a question which has not been systematically addressed before. \rg{While a version of such multi-qubit catalysts has been employed before to accelerate quantum annealing in p-spin models~\cite{Nishimori2017, PhysRevE.85.051112,Nishimori2020}, as we shall elaborate later, the transitions addressed and the mechanism of improvement differ significantly from those discussed here.} 

To do so, we use toy models to characterize first-order phase transitions in the NP-hard maximum weighted independent set (MWIS) problem\cite{Karp1972}. \rgg{Previous studies~\cite{RevModPhys.90.015002,PhysRevA.80.062326} have shown that in traditional adiabatic quantum annealing, the MWIS problem often encounters `perturbative crossings,’ i.e., first-order phase transitions that arise in the regime where the transverse-field terms contribute only perturbatively to the problem Hamiltonian.} Our findings reveal that catalysts entangling multiple qubits are  \textit{crucial} to bypass these phase transitions. We systematically analyze how these catalysts of varying locality can transform phase \textit{transitions into crossovers}\cite{10.1093/acprof:oso/9780199577224.001.0001}, i.e. remove any discontinuities in the energy functional by providing a smooth pathway between the initial state to the desired ground state of the problem.

We also find that sometimes even small $n-$(localized) catalysts suffice, and we explore the underlying reasons. We successfully test our approach on random graphs with MWIS problems, demonstrating the effectiveness of our method in removing first-order phase transitions during quantum annealing. \rg{Additionally, we provide an understanding of why such catalysts are effective in mitigating these transitions, contrasting with the case of the first order transitions in $p-$ spin models.} Finally, we also demonstrate the gate-based implementation of our setup, where we achieve substantial improvements in fidelity by incorporating $n-$local terms. Specifically, we show that to reach a given target fidelity, the introduction of non-local terms using a polynomial number of two-qubit gates leads to an exponential reduction in the total gate count required, \rg{compared to the digitized version of adiabatic quantum annealing using single-qubit quantum fluctuations.} This approach enables more efficient quantum circuit designs, significantly reducing resource requirements for gate based quantum annealing.

\section{Phase transitions in MWIS problem on toy graphs}
\subsection{The MWIS problem and the mapping to Ising model}
The Maximum Weighted Independent Set (MWIS) problem on a weighted graph \( G \) with vertices \( V \) and edges \( E \) is defined as the set of vertices \( V_{M} \) which are not connected to each other by the edges and carry the maximum total weight. The problem can be stated as follows:

Given a graph \( G = (V, E) \) with a weight function \( w: V \rightarrow \mathbb{R} \) assigning a weight to each vertex, the objective is to find a subset \( V_{M} \subseteq V \) such that:
\begin{enumerate}
    \item  No two vertices in \( V_{M} \) are adjacent, i.e., for all \( u, v \in V_{M} \), \( (u, v) \notin E \).\\
    \item  The sum of the weights of the vertices in \( V_{M} \) is maximized, i.e., \( \sum_{v \in V_{M}} w(v) \) is as large as possible.
\end{enumerate}

Formally,
\[ V_{M} = \arg\max_{S \subseteq V} \left\{ \sum_{v \in S} w(v) \ \big| \ \forall u, v \in S, \ (u, v) \notin E \right\}. \]

The MWIS problem is a well-known NP-hard problem in combinatorial optimization, making it a challenging target for both classical and quantum algorithms.
To solve the MWIS problem in a quantum computation setup, the solution is mapped to the ground state of an Ising model with anti-ferromagnetic interactions (see Appendix~\ref{app:MWISising} for details). In this context, the set of vertices with \(\ket{\uparrow}\) states provides the solution. The Ising Hamiltonian is given by:
\begin{equation}
    H_p = \sum_{ij \in E} J_{ij} \sigma^z_i \sigma^z_j + \sum_{i \in V} \left(\sum_{j \in \mathrm{nbr}_i} J_{ij} - 2 w_i \right) \sigma^z_i,
    \label{eq:isingMWIS}
\end{equation}
where \( \sigma^z_i \) are the Pauli-Z operators, \( J_{ij} \) represents the interaction strength between neighboring vertices \(i\) and \(j\),
 \( h_i \) is the external field applied to vertex \(i\),
 \( \mathrm{nbr}_i \) denotes the set of neighbors of vertex \(i\). An additional condition $J_{ij} > {\rm Min}(w_i,w_j)$ is imposed to ensure that the weights do not dominate the `independence' of the two vertices when we compute the ground state. 
 \begin{figure}
     \centering
     \includegraphics[width=0.88\columnwidth]{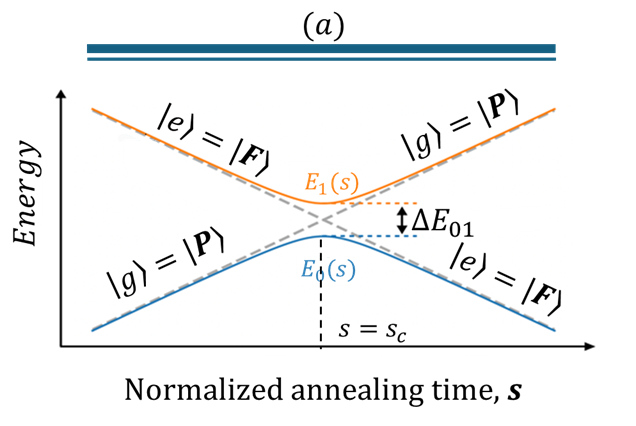}
     \includegraphics[width=0.88\columnwidth]{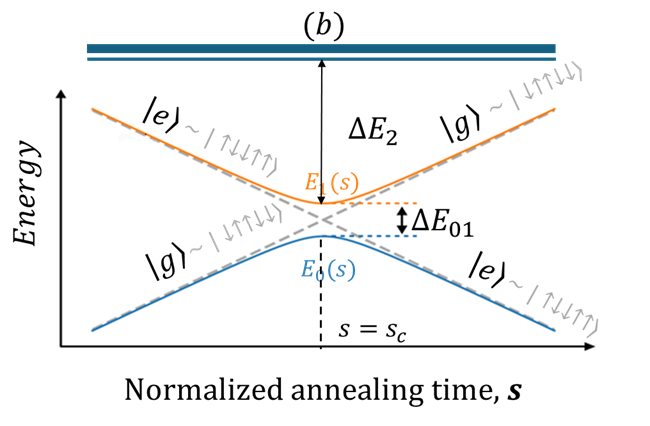}
     \caption{\rg{Schematic diagram of two simple examples of energy gap closures at first order phase transition during quantum annealing. (a) The para-ferro transition where the ground state changes from \rgg{paramagnetic}  $\ket{P}$, to ferromagnetic $\ket{F}$ during the annealing through a first order transition. This 
 occurs for example, in $p-$spin models for odd $p\ge 3$. For even $p$, $\ket{F}$ has a degeneracy. (b) Example of a first order quantum phase transition that can occur deep inside the ordered phase, the `perturbative crossing'. Here the states involved are perturbations around computational states with $O(L)$ Hamming distance between them; an example is provided in the figure. There is a subtle difference in the mechanism of gap closure in the two cases, which is described in detail in the text.}}
     \label{fig:transition}
 \end{figure}
\subsection{Exponential gap closing: Two kinds}
\label{sec:expgap}
\begin{figure}
    \centering
    \includegraphics[width=0.8\columnwidth]{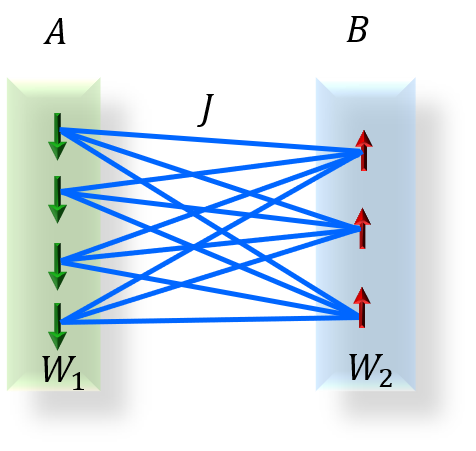}
    \caption{Bipartite toy model, $A$ and $B$ are two subsystems of the bipartite system. We represent the $L=7$ model in the diagram where $A$ has spins $1,2,3,4$ and $B$ has spins $5,6,7$. Partition $A[B]$ has a total weight of $W_1[W_2]$ which is equally divided among the four(three) spins. Thus $w_{1,2,3,4}=W_1/4$ and $w_{5,6,7}=W_2/3$. $J$ is a constant coupling between the spins across the bipartition. \rgg{We typically choose $W_1=W$ and $W_2=W+\delta W$} }
    \label{fig:toy}
\end{figure}

To find the ground state of such problems one can perform adiabatic quantum annealing using the protocol,
\begin{equation}
    H(s)=s H_p+(1-s)H_D
    \label{eq:anneal}
\end{equation}
where $0 \le s \le 1$ is the dimensionless parameter representing annealing time. $H_D$ is the driver Hamiltonian which provides the quantum fluctuations and is given by
\begin{equation}
    H_D=\sum_{i \in V} -\sigma^x_i.
\end{equation}
\rg{The energy of the ground and excited states of $H(s)$ changes throughout the anneal. Two simplest scenarios of occurrence of an energy gap exponentially small in system size \rgg{are} represented in Fig.~\ref{fig:transition}. Let us further assume the energy-gap occurrence is due to a phase transition at $s=s_c$  where the ground and first excited states switch, as shown in the Fig~\ref{fig:transition}. We can write down the Hamiltonian in this two-state subspace at the critical point as
\begin{equation}
    \begin{pmatrix}
   E_0^{(0)}= \langle g| H(s_c)|g \rangle &  \epsilon=\langle g| H(s_c)|e \rangle \\
    \epsilon=\langle e| H(s_c)|g \rangle&
    E_1^{(0)}=\langle e| H(s_c)|e\rangle
    \end{pmatrix},
    \label{eq:matrix}
\end{equation}
where $\ket{g}$ and $\ket{e}$ denote the ground and excited states for $H(s \rightarrow s_c^{-})$, and thus are not eigenstates at $s=s_c$.
The energy gap in this subspace is given by $\sqrt{(E_1^{(0)}-E_0^{(0)})^2+4\epsilon^2}$. The actual energy gap of the problem also contains higher order corrections to this quantity. Now the two distinct ways an exponentially small gap in the spectrum can arise are as follows
\begin{itemize}
    \item For the example in Fig.~\ref{fig:transition}(a), the ground state changes from paramagnetic $\ket{P}=\frac{1}{\sqrt{2^L}}\sum_{i=1}^{2^L}\ket{i}$, where $\ket{i}$ denotes a computational basis state, to the ferromagnetic state $\ket{F}=\ket{1}=\ket{\downarrow\downarrow\downarrow\cdots}$ across the transition, where $L$ denotes the number of sites in the system. Since $H_D=-\sum_{i=1}^L \sigma^x_i$, it is easy to see that $\epsilon\sim O(L/\sqrt{2^L})$, and at criticality $E_0^{(0)}=E_1^{(0)}$~\footnote{The argument also works if $E_1^{(0)}-E_0^{(0)}\sim O(e^{-L})$. The main idea is the approximate or exact degeneracy of the states cannot be lifted by the off-diagonal matrix elements, or quantum fluctuations and hence the gap closes and the system goes through a phase transition. This is best seen in the Grover limit where higher order perturbations are super-exponentially suppressed and an effective $2\times2$ model describes the system.}, hence this energy gap is $2\epsilon$, and it is exponentially small in system size. This is the behaviour seen in $p-$ spin models with $p \ge 3$ and odd. (For even $p$ there is a degeneracy on the ferromagnetic side.) The Grover limit of $p\rightarrow\infty$ also behaves similarly, except here $\epsilon \sim 1/  \sqrt{2^L}$. The higher order corrections can be shown to also be exponentially small~\footnote{For a more rigorous computation of the energy gap see Ref.~\cite{Jorg_2010}}.
    \item Fig.~\ref{fig:transition}(b) illustrates the scenario of primary interest in this work: \emph{perturbative crossings}~\cite{RevModPhys.90.015002}. \rgg{In the problems we discuss, these occur deep within the phase where the ground and first excited states 
typically have maximal overlap with a single, albeit distinct, computational basis state.} We denote these dominant computational states as $\ket{g}$ and $\ket{e}$ respectively, and an example is shown in the figure. Since the eigenstates of $H_p$ are computational basis states, such ground and excited states are perturbative corrections around those eigenstates when we are not at the critical point. Compared to the earlier scenario, these transitions occur when $H_p$ dominates the Hamiltonian, i.e., deep in the `localized' phase.

The distinct feature of this example is that $\epsilon \sim 0$ at the crossing point $s = s_c$, as there exists no direct matrix elements that connect the computational states that dominate the ground and first excited states. They are also separated by $O(L)$ Hamming distance. This aids a principal characteristic seen in a first-order phase transitions: a discontinuous change in an order parameter, often related to magnetization in spin models. During annealing, if the dominant components of the ground and first excited states differ by a large Hamming distance leading to a significant change in magnetization across $s_c$, then typically the system undergoes a first order phase transition. Consequently, since the matrix element between the two states becomes practically zero due to the lack of overlap,  the gap is entirely dependent on higher order corrections.

Using standard high-order degenerate perturbation theory, if the Hamming distance between the states is $n$, then the lowest nonzero contribution comes from terms like $\bra{g}\Pi_{i=1}^n\sigma_i^x\ket{e}$. This implies that the energy gap scales as $\Delta E_{01} \sim O(1/(\Delta E_2)^{n-1})$. Here, $\Delta E_2$ denotes the energy difference between the computational states involved in the perturbative crossing 
and the nearest states in the spectrum connected to them by the driver Hamiltonian $H_D$ (see Fig.~\ref{fig:transition}). 
It effectively plays the role of a barrier height across the local minima in the perturbative regime.
 When $n \sim O(L)$, this results in a gap that scales exponentially with system size~\cite{BAPST2013127}.

Mitigating such perturbative crossings is the central focus of this work. However, in Appendix~\ref{app:p-spin}, we also briefly discuss the possibility of mitigating the earlier type of transition (Fig.~\ref{fig:transition}(a)) using a catalyst, and highlight the qualitative differences in the nature of the catalyst required in each case.
\end{itemize}}
Our \rg{first} objective is to employ toy models to comprehend the fundamental nature of first-order phase transitions with exponentially small energy gap that could arise during annealing in such a setup, and to design suitable catalysts to mitigate them. To begin, we consider a bipartite graph illustrated in Fig.~\ref{fig:toy}. This set up is similar to the one considered in Ref.~\onlinecite{feinstein2023} and can be easily realized in the D-wave Chimera architecture. \cite{boothby2020nextgenerationtopologydwavequantum}. We choose the total number of spins $L=7$ for our numerical tests discussed in the following section, \rgg{and typically choose $W_1=W$ and $W_2=W+\delta W$}.

\rg{To generate the perturbative-crossing-based first-order transitions during the anneal we have to ensure that the two states involved in the transition are a large Hamming distance apart. Due to the perturbative nature of the eigenstates at the crossing, we deduce that this can occur} when the ground and first excited states of \( H_p \) are separated by a large Hamming distance. In the example shown in Fig.~\ref{fig:toy}, this phenomenon arises (see Appendix~\ref{sec:appA}) when \( (W_2 - W_1) < W_2/3 \). This understanding guides us in selecting the parameters for the toy models, which demonstrate a first-order transition during annealing.

Furthermore, the closure of the energy gap during the anneal is attributed to \( H_D \) providing insufficient quantum fluctuations to smoothly drive the system to its ground state, instead causing it to reach a local minimum in the energy landscape with a large potential barrier to the global minimum. \rgg{At the critical point, the ground and first excited states do not get directly coupled by $H_D$, 
owing to the large Hamming distance between their dominant computational basis components. As a result, the minimum gap arises only through high-order ($n$th-order, with $n$ large) processes in perturbation theory, which makes the gap exponentially small.} Notably, if the Hamming distance between the involved states were smaller, the order parameter would not sharply jump, and the gap would not exponentially vanish, as \( H_D \) could connect these states through low-order processes. Note that in case of a highly degenerate subspace, even if the system undergoes a first order phase transition, the gap may be algebraically small due to hybridization opening up alternate pathways to connect the states across the transition~\cite{PhysRevLett.109.030502}. The exponentially closing gaps in perturbative crossing \rgg{of} MWIS problems typically involve ground states with large weights on \rg{$O(L^0)$ computational basis states. In fact for numerically accessible sizes $L\sim O(10)$, we always found involvement of just two dominant computational basis states in the transition. }

These insights suggest the necessity of increasing quantum fluctuations in the system to prevent the exponential gap closing, but not arbitrarily. As we shall see below, in our toy model, only by introducing a `direct-tunnel' coupling between the states involved in the phase transition, which we achieve using a ``product catalyst'' for this model, can we induce level repulsion, thereby transforming the phase transition into a crossover (or the level crossing to a strong anti-crossing). Later on, we shall discuss the limitations of this specific catalyst for generic models and introduce the versatile $n-$local catalyst as a promising alternative.

\section{Using A catalyst to avoid phase transitions}
A catalyst is an additional interaction introduced during the anneal process in Eq.~\eqref{eq:anneal} to alter the anneal trajectory and avoid phase transitions. This interaction is switched off at the beginning and end of the anneal protocol. Thus, a possible annealing trajectory after including the catalyst is given by:
\begin{equation}
    H(s) = s H_p + (1-s) H_D + s(1-s) H_c
    \label{eq:annealcat}
\end{equation}
where \( H_c \) denotes the catalyst Hamiltonian.
\subsection{The product catalyst}
We first introduce a highly non-local catalyst, which we call the product catalyst, that is a product of $\sigma^x$ on all sites. Essentially, this catalyst flips all spins simultaneously, 
\begin{equation}
    H_{cp}=-\prod_{i=1}^L \sigma^x_i,
    \label{eq:prodcat}
\end{equation}
introducing a direct coupling between states separated by a Hamming distance of \( L \), the system size. This prevents any level crossings involving such states.

We verify the efficacy of this catalyst in the toy problem shown in Fig.~\ref{fig:toy}. We first choose an instance of the problem where there is a first order phase transition during the anneal without the catalyst. Subsequently, we include the product catalyst in the anneal protocol and analyze the resulting improvements. We also provide another point of comparison using a different catalyst which is typically seen in literature \cite{PhysRevA.99.042334,PhysRevResearch.3.043013}, the XX catalyst.  The XX catalyst involves adding an $\sigma^x \sigma^x$ interaction to the $\sigma^z \sigma^z$ bonds of the problem, i.e.,
\begin{equation}
    H_{cXX}=-\sum_{ij \in E} \sigma^x_i \sigma^x_j.
    \label{eq: XXcat}
\end{equation}
We shall see later that this forms the simplest example of an $n-$local catalyst. 

\begin{figure}
    \centering

        \includegraphics[width=0.85\columnwidth]{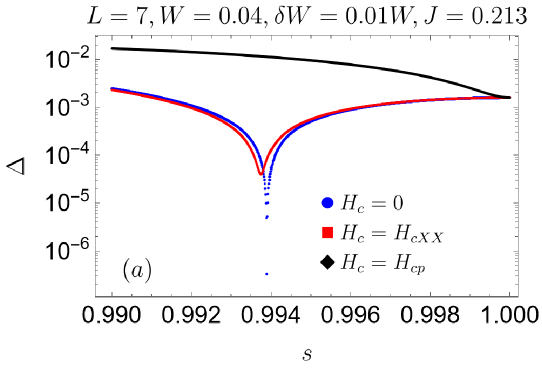}\\
\hspace{0.1 in}

     \includegraphics[width=0.83\columnwidth]{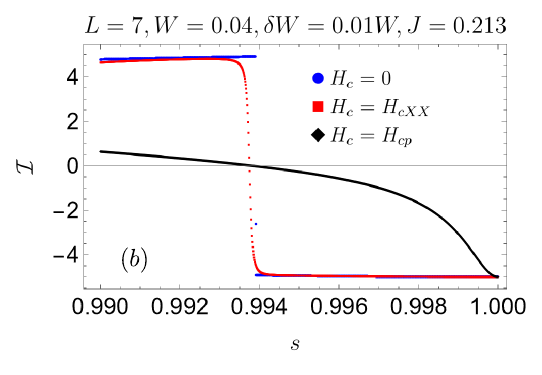}
     \hspace{-0.1 in}
    \caption{(a) The variation of energy gap $\Delta$ with the anneal parameter $s$ for a system of size $L=7$ with $4$ spins in subsystem $A$ and $3$ spins in $B$. The different colours indicate the three different cases. Blue ($H_c=0$) indicates the case when we do not add a catalyst, Eq.~\eqref{eq:anneal}. The others correspond to the anneal according to Eq.~\eqref{eq:annealcat}: red indicates the case where an $XX$ interaction is added on all the $ZZ$ bonds ($H_c=H_{cXX}$) and black represents the case when the product catalyst ($H_c=H_{cp}$) is used. (b) Variation of the order parameter with $s$ showing the presence and absence of a transition in different cases. }
    \label{fig:7site}
\end{figure}

\begin{figure}
    \centering
    \includegraphics[width=0.8\columnwidth]{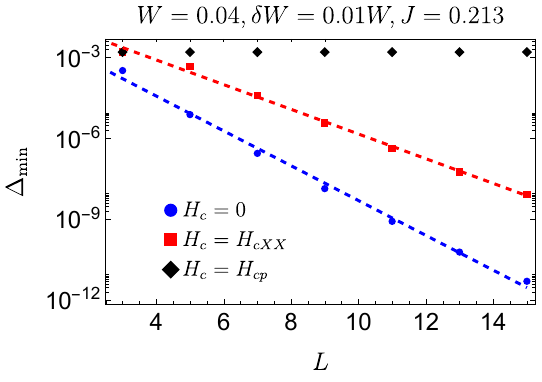}
    \caption{Scaling of minimum energy gap $\Delta_{\rm min}$ with system size $L$ under different catalysts. The dashed line indicates the best fit $\Delta_{\rm min}=A e^{-b L}$. For no catalyst $(H_c=0)$, $b\approx1.54$, and for the multiple XX catalyst $(H_{cXX})$ $b\approx1$.}
    \label{fig:gap}
\end{figure}

In Fig.~\ref{fig:7site} we compare these cases. To identify the first order transition, we show the gap between the ground and first excited states $\Delta$ and also compute an order parameter, the imbalance between the spins in subsystem $A$ and $B$ in this setup, 
\begin{equation}
    \mathcal{I}=\sum_{i \in A} \sigma^z_i-\sum_{i \in B} \sigma^z_i.
    \label{eq:Imbalance}
\end{equation}
We see that the product catalyst, $H_{cp}$, completely removes the transition during the anneal in Fig.~\ref{fig:7site}(a), as the energy gap shows no dip during the anneal and the minimum gap now matches with the problem gap. This is in contrast to the situation when only the driver Hamiltonian is used. Furthermore the order parameter shows a smooth variation with $s$ in Fig.~\ref{fig:7site}(b) on addition of this catalyst. The conversion of the phase transition to a smooth crossover is further corroborated in Fig.~\ref{fig:gap} where we see that on the addition of the product catalyst, the minimum gap $\Delta_{\rm min}$, typically a system-size-dependent object, exactly matches the gap of the problem Hamiltonian. \footnote{ We scale up the system in Fig.~\ref{fig:toy}, by adding a spin to each of the subsystems $A$ and $B$.}

On the other hand, we clearly see that while the XX catalyst, $H_{cXX}$, does increase the gap for the case with total spins $L=7$ as shown in Fig.~\ref{fig:7site}(a), the order parameter still shows a sharp jump at the critical point in Fig.~\ref{fig:7site}(b) showing that the transition persists. This is corroborated when we study the scaling of the minimum gap, $\Delta_{\rm min}$ in Fig.~\ref{fig:gap}, which  still shows an exponential scaling with system size $L$. Evidently, the multiple XX catalyst improves the pre-factor of exponential in the gap scaling, from $\sim3/2$ to $\sim1$ in this case,  but has not actually changed the order of the phase transition.  

\rg{Let us briefly comment on the role of the \textit{sign} of the catalyst couplings, in light of earlier suggestions that \textit{non-stoquastic} catalysts may be essential to achieving quantum advantage in certain cases~\cite{Seki2012, Seoane_2012_2, crosson2014differentstrategiesoptimizationusing, Seki_2015, Nishimori2017_2, PhysRevB.95.184416}. In the example we study here, we find that the impact of the catalyst's sign depends strongly on its structure.

Specifically, for the \textbf{XX-catalyst} $H_{cXX}$, changing its sign---i.e., making it non-stoquastic---actually reduces the minimum gap, thereby eliminating any potential advantage. In contrast, for the product catalyst, $H_{cp}$, changing the sign has \textit{no effect} on the gap (see Fig.~\ref{fig:nonstoq}(c) in Appendix~\ref{app:appB}).

This contrast can be understood as follows: the two states involved in the transition are separated by the maximal Hamming distance $L$. Without any catalyst, the off-diagonal coupling $\epsilon$ in Eq.~\eqref{eq:matrix} vanishes. The product catalyst directly connects these two states, introducing a nonzero matrix element $\epsilon \sim s_c(1 - s_c)$, which hybridizes them. This is a \textit{non-perturbative} effect---the corrections due to $H_D$ appear only at higher order and are negligible (as confirmed numerically for a toy model). Moreover, the effective gap scales as $\sim \sqrt{4 \epsilon^2}$, so changing $\epsilon$ to $-\epsilon$ has no impact on the gap size.

By contrast, the XX catalyst does not directly couple the two states due to their large Hamming separation. If we denote a basis state with Hamming weight $0 \le h \le L$ as $\ket{h}$, the only nonzero matrix elements of $H_{cXX}$ are of the form $\bra{h} H_c \ket{h \pm 2}$. Thus, hybridization must proceed via virtual transitions through intermediate states, making it a higher-order, perturbative process. In such cases, the sign of the catalyst influences how these terms interfere, and its effect becomes highly dependent on the problem structure.

We explain the role of perturbative catalysts in greater detail in the following section, and also discuss why non-stoquastic catalysts are particularly effective in mitigating the ferro-para first-order transition in contrast to their behaviour in the models considered here, in Appendix~\ref{app:p-spin}.
}

However, while the toy example serves as an excellent demonstration of the necessity of the non-local nature of catalysts in removing the phase transition, the product catalyst works only for this specific problem structure. For other problem structures one would need to figure out the states involved in the possible first order phase transition to develop the catalyst that removes them. This typically is a hard problem  \cite{Takahashi_2019,PhysRevResearch.5.043236}, and hence we  utilize the essence of the product catalyst, the multi-qubit coupling, to develop other catalysts which can work in more general scenarios. \rg{Let us first motivate why this can work.

Perturbative crossings arise when the ground and excited states are dominated by computational basis states that differ by a large Hamming distance \( h \). In such cases, the standard driver \( H_D \) can only couple them at the \( h^{\text{th}} \) order in perturbation theory, leading to an exponentially small gap when \( h \sim L \). Then, naively, introducing an \( n \)-qubit catalyst can allow these states to be coupled at lower order, approximately \( h/n \), thereby enhancing level repulsion and alleviating the exponential gap suppression. In the next section, we do a quantitative analysis of this idea using numerical examples and delve deeper into the mechanism of opening up the energy gap using such catalysts. We shall see that the perturbative effects described for the $XX$ catalyst in this section will play a crucial role in our understanding.
}
\subsection{$n-$local catalysts}
\begin{figure}
    \centering
    \includegraphics[width=0.8\columnwidth]{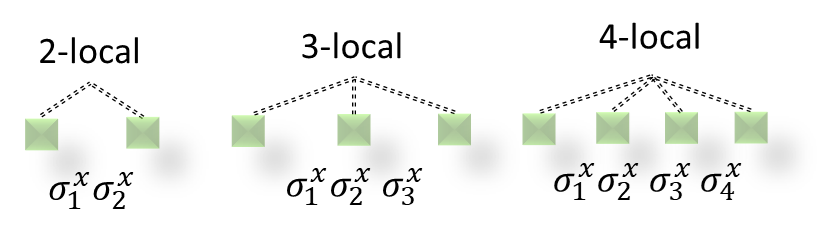}
    \caption{Examples of $n-$local catalysts with n$=2,3,4$. The simplest is 2-local, which corresponds to the buliding blocks of $H_{cXX}$ in Fig.~\ref{fig:7site}. The product catalyst is the extreme case of $n-$locality where all the sites are connected. }
    \label{fig:n-local}
\end{figure}
While the product catalyst induces a non-perturbative transformation of the phase transition to a crossover, it is effective only when it connects the corresponding states across the transition. This indicates that merely maximizing the strength and range of quantum fluctuations does not guarantee quantum advantage; only a direct-tunnel coupling can eliminate phase transitions. In most geometries, not all spins change their orientation during the phase transition, rendering the product catalyst ineffective. Additionally, predicting which spins are involved in the transition a priori is nearly impossible with just the problem statement. However, the key takeaway from the product catalyst is the necessity of introducing entangling quantum fluctuations that connect states which are separated by large Hamming distances. In essence, this can also be achieved by using $n-$local catalysts which flip $n$ spins together, instead of all of them as in the product catalyst. In optimal cases, this either enables tunneling between relevant states across the transition or selects an anneal pathway that avoids the transition. Even in the worst-case scenarios, increased quantum fluctuations is typically expected to increase the energy gap across a transition, thus improving annealing time\cite{PhysRevA.95.042321}. \rg{Note however one needs to be careful in controlling the strength of the $n-$local catalyst since the number of couplings increases exponentially as $n$ increases. We rescale the strength of the catalyst, if necessary, so as to ensure at all orders of $n-$locality, $||H_c||$ does not exceed  $||H_P||$.}

In Fig.~\ref{fig:n-local} we illustrate the concept of $n-$local catalysts.
\begin{equation}
   H_{cn} =-\prod_{i=1}^n \sigma^x_i
    \label{eq:nlocalcat}
\end{equation}
\rgg{Introducing additional quantum entanglement between $n$ sites during the anneal, 
this catalyst effectively smooths the potential landscape in a way that is broadly applicable 
across different random graph structures, thereby providing a more structure-agnostic approach to accelerate adiabatic quantum computing.} In what follows, we shall show that this catalyst is very effective for the MWIS problem defined in Eq.~\eqref{eq:isingMWIS}. In fact, we shall deduce that its performance varies based on the sites which the catalyst connects.  In some problem instances, correctly arranging even the $2$-local catalysts can eliminate the transition altogether, thus achieving exponential improvement in gap scaling. Note that we shall only consider optimization of arrangements of couplings in this work and not the strength of the catalyst for a simplified analysis. Again, we shall first analyze the toy problem to understand the working principles.

\begin{figure*}
    \centering
    \includegraphics[width=0.85\columnwidth]{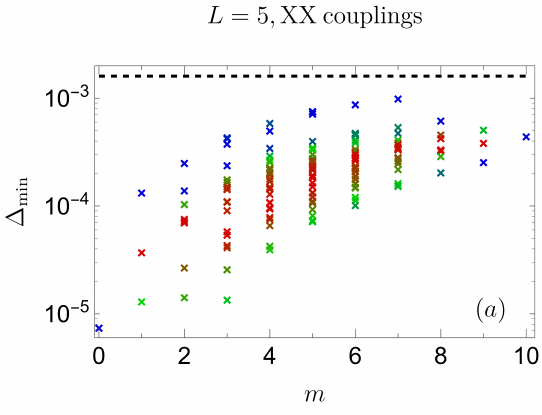}
      \includegraphics[width=0.85\columnwidth]{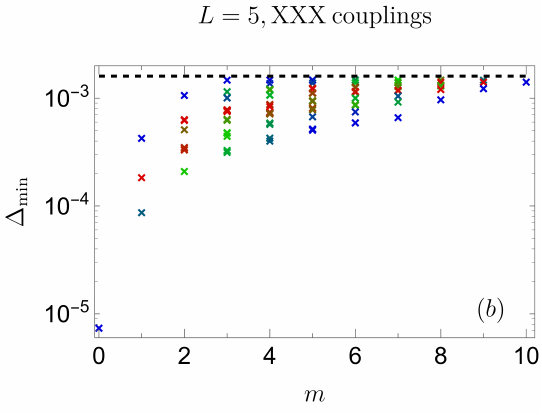}
         \includegraphics[width=0.85\columnwidth,height=0.7 \columnwidth]{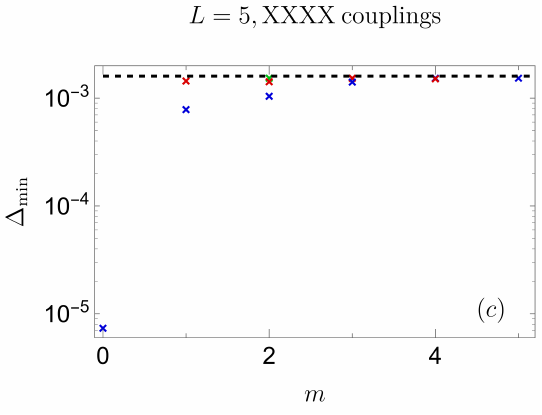}
            \includegraphics[width=0.83\columnwidth,]{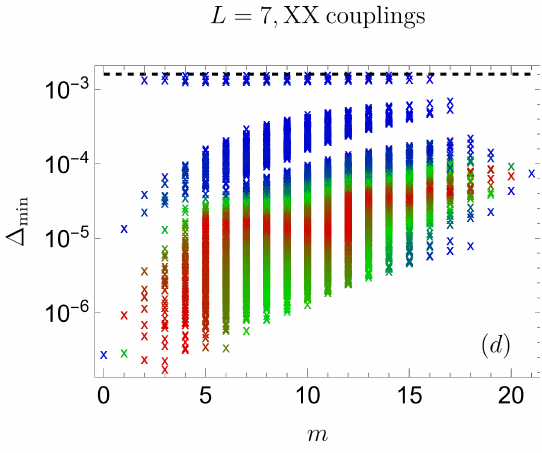}
    \caption{(a),(b),(c) Plots showing various scenarios of using $2-$local (XX), $3-$local (XXX) and $4-$local (XXXX) couplings for L=5. $m$ denotes the number of couplings of the catalyst and each point denotes a different configuration of the $m$ couplings. The black dashed line shows the energy gap of the problem, $H_p$. Reddish hues indicate a high density of points, \rg{blue indicates a low density and green indicates intermediate densities}. (d) shows the XX catalyst for $L=7$.  (a) and (b) feature $10^3$ configurations each, (c) features $10$ configurations and (d) features $2.097 \times 10^6$ configurations.}
    \label{fig:5siteoptimize}
\end{figure*}
\paragraph{\textbf{Revisiting the toy problem:}} In the toy example of Fig.~\ref{fig:toy}, we shall now add $n-$local catalysts instead of a single product catalyst to assess potential improvements. There are approximately \( 2^{L \choose n} \) ways to add such catalysts due to the choices of number and placement of catalysts, which becomes intractable for \( n > 3 \) and \( L \ge 7 \) (For $n=2$ and $L=7$ we have $2 \times 10^6$ possible combinations.)  In Fig.~\ref{fig:5siteoptimize}(a),(b) and (c), we plot the minimum energy gap during the anneal, \( \Delta _{\rm min}\), against a fixed total number $m$ of $n-$local interactions added for \( n = 2, 3, 4 \) (respectively) for $L=5$. The different data points exhaust all possible configurations of the $n-$site interactions.\footnote{Due to the problem geometry, some configurations are equivalent, but we do not make a distinction here.} For example, for the XX catalyst, $3$ on the x-axis denotes the scenario where we add the interactions between any three pairs of vertices, which may or may not include the edges of the problem statement. We plot all possible arrangements of such a catalyst for $m=3$ as different data points. 
Additionally, we also present the result for the only tractable case for $L=7$ in Fig.~\ref{fig:5siteoptimize}(d) for the $2-$local catalyst.  

Our findings align with the intuition that larger \( n \) in $n-$local catalysts enhances the energy gap improvement in such geometries. Furthermore, on average, the more vertices connected by interactions, the better their performance. However, interestingly, if positions of the catalysts are optimized correctly, even 2-local catalysts can be nearly as effective as the product catalyst in closing the energy gap. Nevertheless, these configurations are rare, and the majority of configurations yield significantly less gap increase than the product catalyst. The mean improvement in each case approaches the product catalyst's efficacy as both the number and the value of \( n \) increase.

\begin{figure}
    \centering
    \includegraphics[width=0.48\columnwidth]{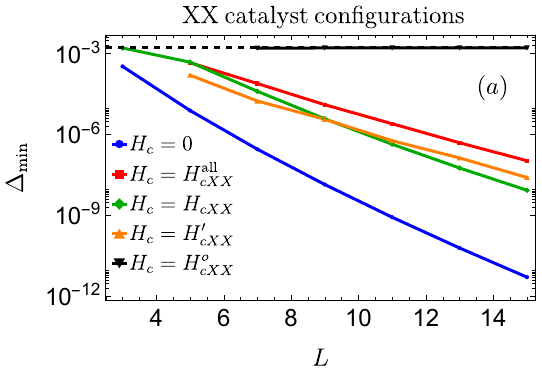}
        \includegraphics[width=0.48\columnwidth]{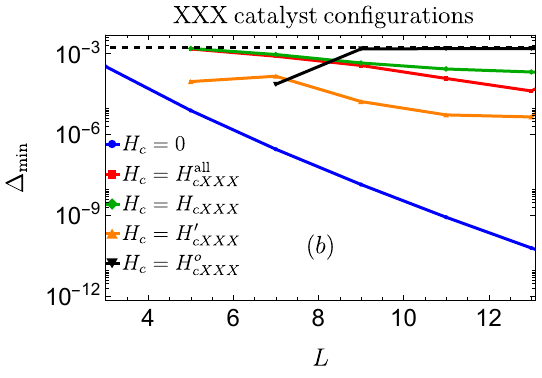}
    \caption{Scaling of the minimum energy gap $\Delta_{\rm min}$ with system size for a few chosen configurations of the catalyst. We compare the case with no catalyst $H_c=0$, to adding a catalyst on all possible coupling sites (superscript `${\rm all}$'), adding a catalyst only on the problem edges ($H_{cXX}$ and $H_{c XXX}$), adding a catalyst avoiding the problem edges ( $\prime$)  and an optimal configuration (superscript $o$) for (a) the $2-$ local XX catalysts and (b) the $3-$ local XXX catalysts. The lines are a guide to the eye. The optimal catalyst configuration $H^o_{cXX}$ has the same number ($m=3$) and configuration of couplings for different system sizes $L$ showing consistent improvement irrespective of the size of the problem. For $H_{cXXX}^o$, $m=2$ throughout. }

    \label{fig:cateffect}
\end{figure}

\paragraph{\textbf{Scaling of gap for different catalyst configurations with system size}:} \rgg{We look into this phenomenon further in Fig.~\ref{fig:cateffect}, where we study the effect of system size on specific catalysts of $n-$locality of $2$ and $3$. We scale up the toy model in Fig.~\ref{fig:toy} to do so. The several catalyst configurations we compare are}: (i) adding a coupling on all possible two-site ($H_{cXX}^{{\rm all}}$) and three-site combinations ($H_{cXXX}^{{\rm all}}$) i.e., adding \({L \choose 2}\) and \({L \choose 3}\) couplings respectively, (ii) coupling all sets of two and three spins with edges between them in the problem statement ($H_{cXX}$ and $H_{cXXX}$), (iii) coupling all sets of two and three spins that have no edges between them ($H_{cXX}^{\prime}$ and $H_{cXXX}^{\prime}$), and (iv) an optimal catalyst configuration ($H_{cXX}^{o}$ and $H_{cXXX}^{o}$ ) in the toy problem of Fig.~\ref{fig:toy}. These choices constitute some natural ways one can arrange the catalyst couplings in a generic problem. \rg{The optimal catalyst configuration, determined at small $L$, remains unchanged as $L$ increases. That is, the same arrangement and number of couplings apply for all $L$ values shown. Specifically, only $3$ couplings are needed for the $XX$ case and $2$ for the $XXX$ case.} Overall, we observe that barring the optimal catalyst, for $2-$local XX catalysts, $H_{cXX}^{{\rm all}}$ yields slightly better results than the rest \footnote{To achieve this, one typically needs to embed the problem into a unit disk graph, which leads to redundant qubits. Therefore, in practice, adding XX catalyst to the edges might be preferable.}. \rgg{Further improvement is observed when employing the $XXX$ catalysts, with $H_{c}^{XXX}$ yielding the most pronounced enhancement. 
We now provide an analytical investigation of the mechanism underlying this gap increase and the emergence of optimal catalyst configurations}.

\paragraph{\textbf{Mechanism of gap increase:}} We can obtain a qualitative understanding of how different catalyst couplings affect the energy gap by analyzing their perturbative effect on the bare eigenstates of the standard quantum annealer. \rg{Note that we compute the change in the instantaneous ground and excited state energies of $H(s)$ to assess the effect of the catalyst on the instantaneous gap. Since we aim to mitigate perturbative crossings, perturbation theory offers valuable insight into these changes.}
Let us first recall how energies are affected on application of a perturbation. If $\lambda V$ is the perturbation matrix and $E_n^{(0)}$ and $\ket{\psi_n^{(0)}}$ are the unperturbed energy and eigenfunctions, then the perturbative correction to energy up to third order is,

\begin{equation}
\begin{aligned}
E_n \approx & \, E_n^{(0)} + \lambda \langle \psi_n^{(0)} | V | \psi_n^{(0)} \rangle \\
& + \lambda^2 \sum_{m \neq n} \frac{|\langle \psi_m^{(0)} | V | \psi_n^{(0)} \rangle|^2}{E_n^{(0)} - E_m^{(0)}} \\
& + \lambda^3 \left( \sum_{m \neq n} \sum_{k \neq n} \frac{\langle \psi_n^{(0)} | V | \psi_m^{(0)} \rangle \langle \psi_m^{(0)} | V | \psi_k^{(0)} \rangle \langle \psi_k^{(0)} | V | \psi_n^{(0)} \rangle}{(E_n^{(0)} - E_m^{(0)})(E_n^{(0)} - E_k^{(0)})} \right. \\
& \left. - \langle \psi_n^{(0)} | V | \psi_n^{(0)} \rangle \sum_{m \neq n} \frac{|\langle \psi_m^{(0)} | V | \psi_n^{(0)} \rangle|^2}{(E_n^{(0)} - E_m^{(0)})^2} \right).
\end{aligned}
\end{equation}

For corrections to the ground state, note that $E_m^{(0)} > E_0^{(0)}$ for all $m > 0$, so $E_n^{(0)}-E_m^{(0)}$ in all terms are negative. \rg{If the catalyst has just $\sigma^x$ terms and is stoquastic, i.e. ${\rm sgn}(V) = -1$, then by the Perron–Frobenius theorem, the first-order term (if non-zero) is negative, lowering the ground state energy}. \rg{Additionally, for the second-order term}, the numerator is positive and the denominator negative, yielding another negative correction. \rgg{For the third-order term, when the first-order contribution vanishes 
(as often happens when the ground state has dominant overlap with a single computational basis state), 
and assuming non-degenerate classical solutions, 
the dominant contributions arise from low-lying states that also have large overlap with single computational-basis state, say $\ket{s_k}$, i.e. \[
|\psi_k^{(0)}\rangle \approx |s_k\rangle + \mathcal{O}(\epsilon).
\]
Since $V$ is stoquastic in $\ket{s_k}$ basis, the corresponding matrix elements carry negative signs, 
so the resulting numerator has an opposite sign to the denominator, lowering its energy. 
Possible contributions from higher excited states may have different signs, 
but these are suppressed by their much larger energy denominators in the perturbation expansion.
} Thus, stoquastic catalysts perturbatively lower the ground state energy in the first few orders \rgg{in the regime of perturbative crossings in problems with non-degenerate solutions}.~\footnote{ \rg{Although higher-order terms can have mixed signs, many-body perturbation theory is asymptotic and non-convergent. We therefore focus on low-order corrections assuming they dominate.}}

Now if such a catalyst strongly hybridizes the ground state with a higher energy state, which is energetically close to the ground state ($E_m^{(0)}-E_0^{(0)} \gtrsim |\langle \psi_m^{(0)} | V | \psi_0^{(0)} \rangle|$), there will be a significant reduction in the ground state energy. If the same catalyst connects the first excited state only weakly to  all other states ($|E_m^{(0)}-E_1^{(0)} |\gg |\langle \psi_m^{(0)} | V | \psi_1^{(0)} \rangle|\: \forall m $), then the perturbation effectively `pulls'  one state apart from another and opens the gap~\footnote{If the ground state and first excited state are the two states involved in the strongest hybridization,  then we go back to the direct-coupling scenario. However even from this theory it should be evident that a direct coupling affects the ground and first excited state in the opposite manner and pulls them apart.}. This is the underlying mechanism behind the optimal catalyst configurations. Note that if the catalyst is non-stoquastic then, in this perturbative limit, the first few terms in the series will instead be of opposite signs and the improvements much more difficult to assess  (see also Fig.~\ref{fig:nonstoq} in Appendix.~\ref{app:appB}). However, as shown before, this distinction goes away in the non-perturbative limit when several computational states \rg{directly} hybridize with each other.

Depending on the problem statement (moving away from our toy model) the state energetically close to the ground state may be a large Hamming distance away. If we can directly hybridize these two states strongly, the energy gap opened will nearly match or exceed that of the gap of the problem statement and we will achieve the optimal scenario. The product catalyst and certain optimal $n-$local coupling configurations in our toy model achieve this effect. Note also that if a catalyst connects too many states, its chances to affect both the ground and excited states increases and optimal configurations may not be found. This is clearly evident in Fig.~\ref{fig:5siteoptimize}(a) and (d), where no optimal catalyst configurations exist when large number of catalysts couplings ($m$) are added. \rgg{However, this feature is less apparent in Fig.~\ref{fig:5siteoptimize}(b) and (c) because, for small systems where the catalyst locality $n$ is comparable to the system size $L$, 
the catalyst directly hybridizes most computational states through low-order processes, 
and the perturbative gap scales as
$
\Delta_{\min} \sim \frac{1}{(\Delta E_2)^{O(1)}},
$
resulting in uniformly opened gaps across the spectrum. 
The $XXX$ and $XXXX$ couplings for $L=5$ fall into this regime. 
When $n \ll L$, however, connecting distant configurations requires higher-order processes of order $O(L/n)$, 
so only specific catalyst structures that couple the relevant low-lying states 
can effectively enhance the gap, which is observed for the $XX$ couplings even at these system sizes. 
The system-size scaling shown in Fig.~\ref{fig:cateffect} corroborates this interpretation. 
This understanding becomes particularly important when extending our method to random graph structures.
} 

\rgg{For arbitrary random geometries, a brute-force search for the optimal configuration of catalyst couplings 
rapidly becomes intractable, as there are $O(2^{L^a})$ possible combinations with $a \ge 2$. 
It is therefore necessary to restrict attention to a few generic but physically motivated catalyst configurations. 
Insights from the toy model suggest that adding couplings along the edges of the problem graph provides a reasonable baseline strategy. 
At the same time, as indicated by the preceding analysis, increasing the locality $n$ of the catalyst tends to enhance performance for this class of problems. 
Hence, designing effective yet scalable configurations of $n$-local catalysts offers a practical and promising route to improving annealing performance in random graph instances.}

\begin{figure}
    \centering
    \includegraphics[width=0.48\columnwidth]{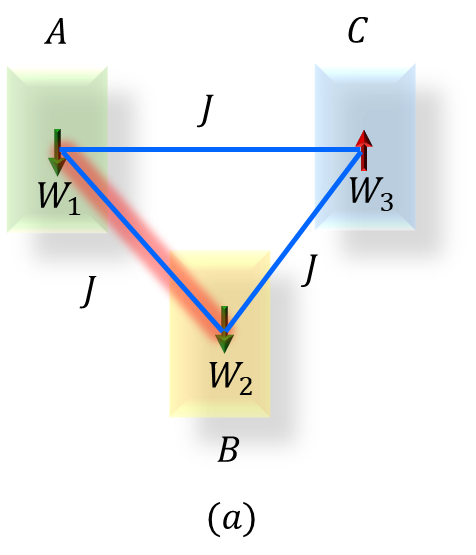} \\
\hspace{0.05 in}      \includegraphics[width=0.78\columnwidth]{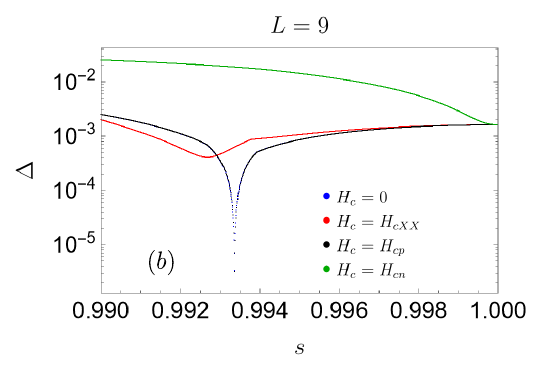}
    \caption{(a)Schematic diagram of a tripartite setup. The bond with a red background is the frustrated bond. (b) Numerical data verifying that the product catalyst does not work in the tripartite scenario. The tripartition has been made to $2,3$ and $4$ spins. $[W_1,W_2,W_3]=[W/2,(W-\delta W)/3, (W-2 \delta W)/4]$ where $W=0.04, \delta W=0.01 W$ and $J=5.33 W$. $H_{cn}$ {=} $-\Pi_{i \in A, B} \sigma^x_i-\Pi_{i \in B, C} \sigma^x_i-\Pi_{i \in A, C} \sigma^x_i$}
    \label{fig:tripartite}
\end{figure}

\paragraph{\textbf{Identifying redundant couplings:}}One way to make an informed choice of a catalyst is to identify structures that fail to increase the energy gap for specific optimization problems, thereby reducing the pool of choices. Fortunately, for the MWIS problem, we have a way of doing so. In what follows we shall show that no gap enhancement occurs when an $n-$local catalyst is connected across a frustrated loop of the graph for such problems. This is because such a catalyst cannot eliminate any first-order phase transition across the loop. To illustrate this, let us consider the problem structure in Fig.~\ref{fig:tripartite}(a) which represents a tripartite lattice with a frustrated bond.
The frustration exists due to the antiferromagnetic nature of the interaction, which results in all three bonds being unable to be satisfied. Note that in Fig.~\ref{fig:tripartite}(a) we have shown just one spin in each block in the figure for simplifying the analysis;  the following arguments can be easily generalized to a set of spins in each block.
Let us also assume we have  $(W_1,W_2,W_3)$ as the three weights and $J_{12},J_{23},J_{13}=J$ as the three couplings for simplicity. Following the MWIS problem structure, without loss of generality we have two cases,
\begin{itemize}
    \item $J>W_3>W_2>W_1$: The ground state for this setup is shown in Fig.~\ref{fig:tripartite}(a). In this case, flipping spin $A$ or $B$ costs $4J - 4W_{1,2}$, while flipping spin $C$ costs $4W_3$. Flipping both $B$ and $C$ spins together costs $4(W_3 - W_2)$, flipping $A$ and $C$ spins together costs $4(W_3 - W_1)$, and flipping all three spins costs $4J - 4W_1 - 4W_2 + 4W_3$. From the problem statement, since $J > W_i$, we can see that flipping all three spins costs more than flipping two spins, which in turn costs less than flipping one spin in this scenario. Thus, the first excited state is given by a state at a Hamming distance of 2 from the ground state. Note that if all three bond strengths were not the same, flipping all spins would cost $4J_{13} - 4W_1 - 4W_2 + 4W_3$, and the same argument would still hold.

    \item $W_3>J>W_2>W_1$: The ground state remains the same in this case. However, due to the change in the statement of the problem, we now have $J - W_{1,2} < W_3 - W_{1,2}$, which indicates that the single spin flip is the first excited state. The all-flipped state remains energetically unfavorable, and the frustrated bond in the system ensures this outcome in both scenarios.

\end{itemize}
As discussed earlier, the states involved in the phase transition are typically the lowest energy states of the problem. Clearly, the all-spin-flipped state is not a part of them under any allowed circumstances. This means that the product catalyst will fail to improve the energy gap for this geometry as it is \textbf{not} the direct-tunnel coupling catalyst. In Fig.~\ref{fig:tripartite}(b), we provide numerical evidence of this statement. We show the behavior of a tripartite system with a phase transition during the anneal under the presence of both the product catalyst and a tailor-made $n-$local catalyst for the geometry. In this case, the optimal $n-$local catalyst involves connecting all the spins in two blocks at a time. This is because the chosen parameters correspond to the first case discussed above but with multiple spins in each block. Thus, the first excited state involves flipping two blocks of spins. Hence, an $n-$local catalyst successfully removes the phase transition, whereas the product catalyst does not.

However, since the $n-$local catalyst does not connect two states that are farthest apart in Hamming distance, it will have perturbative effects on states at a Hamming distance greater than \(n\). Consequently, the sign of the catalyst interaction becomes significant. We have verified (see Appendix~\ref{app:appB}) that the stoquastic choice is preferable, for the same reasons discussed previously. 
Note that $H_{cXX}$ also shows perturbative improvement in the energy gap, as expected, but does not remove the phase transition altogether.

\begin{figure}
    \centering
    \includegraphics[width=0.75\linewidth]{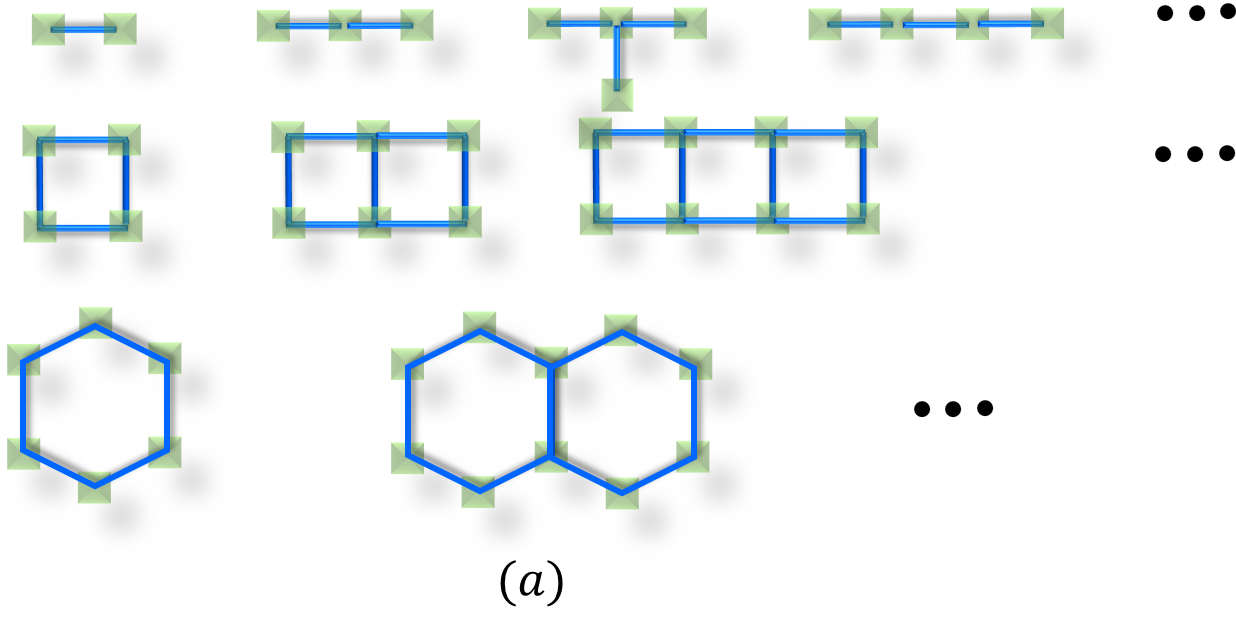}\\ \hspace{-0.2in}
    \includegraphics[width=0.33\linewidth]{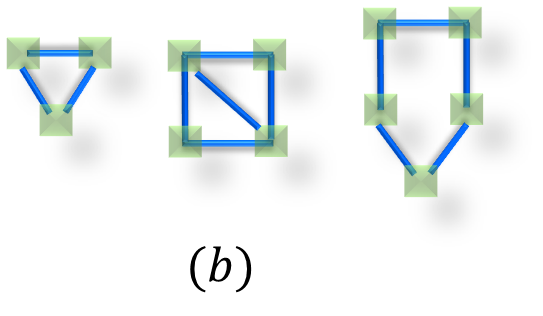}
    \caption{(a) Examples of subgraphs with low $n$ where $n-$local catalysts should be connected across all vertices to obtain speed up. The blue lines denote edges in subgraphs of the problem graph and the green squares denote the vertices in the problem graph. The first row represents all the tree subgraphs. (b) Examples of loops in which connection of $n-$local catalysts (where $n$ equals number of vertices) {\bf does not} yield improvement. Thus one needs to connect the $n-$local catalyst across only those loops whose all subloops have even number of vertices.}
    \label{fig:possible cases}
\end{figure}

\begin{figure}
    \centering
    \includegraphics[width=0.5\linewidth]{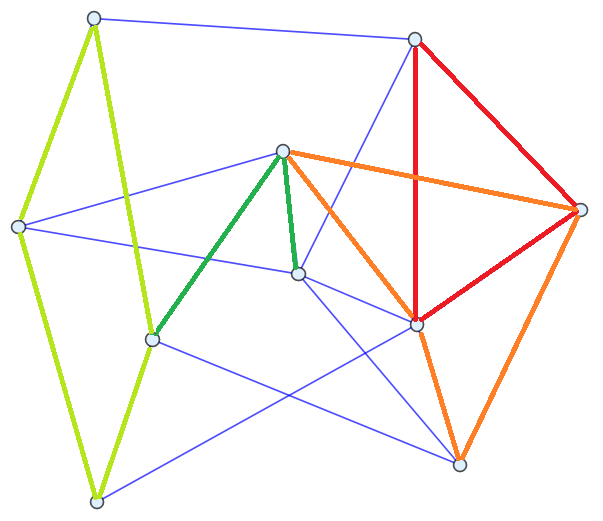}
    \caption{An abitrary graph showing the subgraphs across which adding a catalyst is beneficial (thicker lines in shades of green), and useless (shades of red). An example of a beneficial $3-$local catalyst is shown in dark green and a beneficial $4-$local catalyst is shown in light green, whereas an example of a useless $3-$local catalyst is shown in red, and a useless $4-$local catalyst in orange.}
    \label{fig:examplesubgraph}
\end{figure}

This understanding extends to all types of frustrated loops with an odd number of nodes. In such cases, applying quantum fluctuations by connecting an $n$-local catalyst across all nodes of the loop is ineffective for MWIS problems. \rg{This can be demonstrated through a simple picture. In such a loop, the maximum number of $\ket{\uparrow}$ spins allowed in the ground state is $(L - 1)/2$, where $L$ is the number of vertices, to preserve the independence of the solution set. Flipping all the spins in the loop results in a configuration that is no longer an independent set, as it necessarily includes at least one pair of adjacent $\ket{\uparrow}$ spins. This effect is what is illustrated in the simplest case in the tripartite scenario discussed above. The resulting configuration has a much higher energy due to the structure of the MWIS problem, where $J_{ij} > \min(w_i, w_j)$. This is because now we have $(L+1)/2$ \ket{\uparrow} spins ($L=3$ in the above example) and this increases the energy by $ O(4J)$ from the on-site $\sigma^i_z$ term, and thus no longer remains a possible low energy excited state. Consequently, connecting this state to the true ground state is ineffective, as the effective hybridization of the ground state and the all spin flipped state would be too negligible to significantly change the ground state energy.} This allows us to outline a possible hierarchy of $n-$local catalysts that can be added to subgraphs of arbitrary graphs to improve the energy gap. \rgg{The basic guiding principle is that they should not have any frustrated loop connections}, as shown in Fig.~\ref{fig:possible cases}(a). On the other hand in Fig.~\ref{fig:possible cases}(b) we show examples of possible subgraphs where adding the catalyst offers no improvement. The key observation from Figs.~\ref{fig:possible cases}(a) and (b) is that loops with $n$ odd, i.e., having an odd number of vertices, should not be connected by a coupling of locality $n$ or greater. \rg{For further clarity of understanding, in Fig.~\ref{fig:examplesubgraph}, we illustrate examples of advantageous subgraphs (shaded in green) where applying an $n$-local catalyst across the participating vertices can lead to improved performance.} In shades of red, we also show examples of those subgraphs across which connecting an $n-$local catalyst will offer no improvement.

Hence, to improve the performance of a quantum annealer, one should always first introduce $2$-local couplings, which will perturbatively improve the energy gap in many scenarios \cite{nutricati2024enhancingenergygaprandom}. But to achieve further improvement, one should introduce $3$-local couplings that do not involve connecting triangular loops. Then, if further improvement is sought, different $4$-local couplings can be introduced, but in this case, only square loops which contain no triangular loops should be connected. This hierarchy can continue until the limitations of the experimental setup are reached, or until no improvement is observed in the outcomes.

\begin{figure}
    \centering
    \includegraphics[width=0.65\columnwidth]{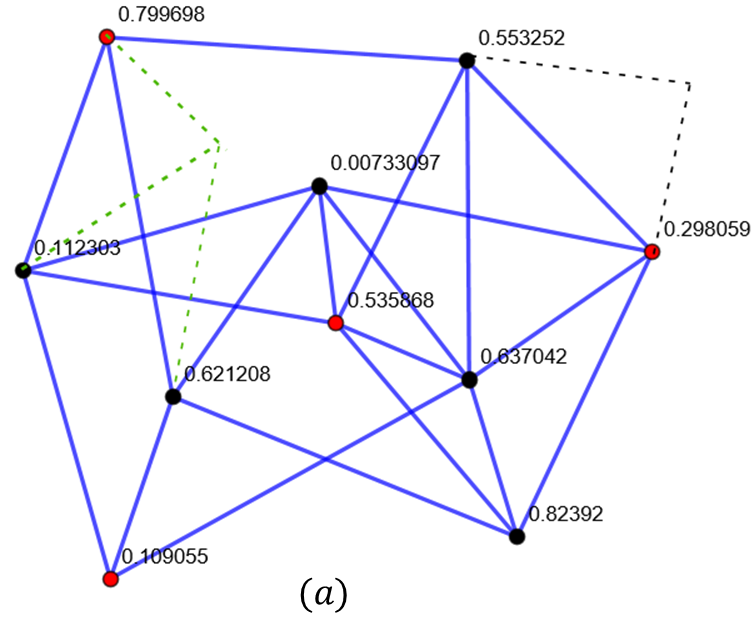}
    \includegraphics[width=0.85\columnwidth]{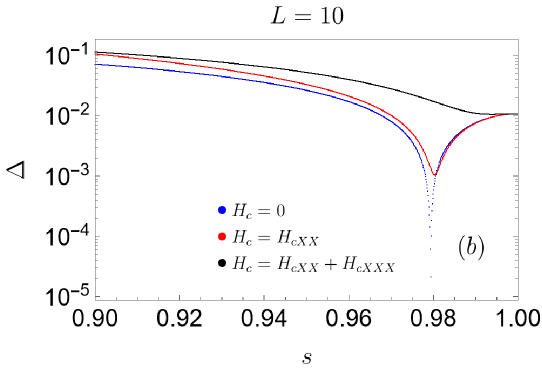}
    \caption{(a) An example of a random graph where adding the XXX couplings removes the phase transition but the $XX$ couplings alone cannot. The dashed lines indicate two examples of the many couplings present in the catalyst. The red dots indicate the sites which provide the solution to the problem instance. The couplings are added to all the allowed nodes, i.e restricted to the first two scenarios of Fig.~\ref{fig:possible cases}. Notably, while $H_{cXXX}$ should naively include $21$ additional couplings compared to the $H_{cXX}$ case, removing the frustrated loops reduces $33\%$ of the couplings, and $14$ couplings are sufficient to obtain the improvement. The numbers on each vertex indicate the weights $w_i$. (b) Variation of energy gap $\Delta$ with the anneal parameter $s$ demonstrating how the addition of $H_{cXXX}$ removes the gap closure at the phase transition.}
    \label{fig:XXXbetter}
\end{figure}

\paragraph{\textbf{An example in a random graph:}} In Fig.~\ref{fig:XXXbetter} we show an example of a random graph, where adding a 3-local catalyst ($H_{cXXX}$) following the hierarchy helps to increase the overall energy gap as well as remove the first order phase transition, while just using $H_{cXX}$ do not\cite{nutricati2024enhancingenergygaprandom}. The graph is generated as an Erdős-Rényi graph with the probability of an edge between two vertices $p=0.5$. We choose an instance which exhibits a phase transition during the anneal. The weights are chosen randomly from $[0,1]$ and are shown on the vertices of Fig.~\ref{fig:XXXbetter}(a). The bond strengths are also randomly chosen from a uniform distribution $[1,2]$, not shown in the figure to avoid cluttering. We provide details on the bond strengths in Appendix~\ref{app:appC}. 

\begin{figure*}
    \centering
\includegraphics[width=0.66\columnwidth]{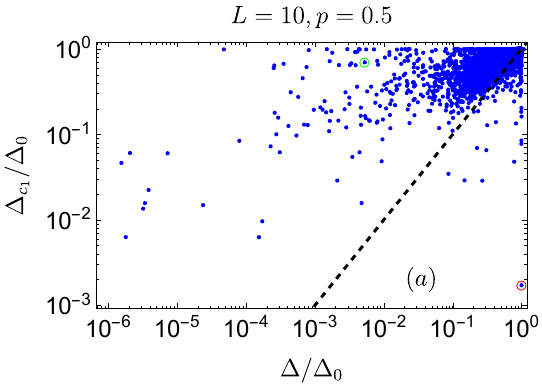}
\includegraphics[width=0.66\columnwidth]{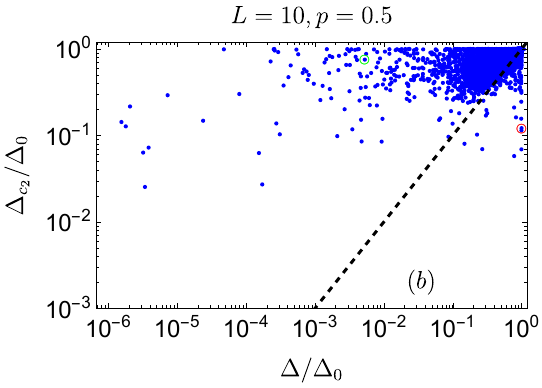}
\includegraphics[width=0.66\columnwidth]{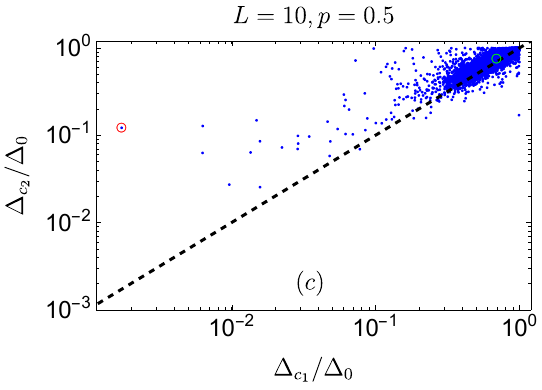}
\caption{Plot showing the improved gap opening for XXX catalysts vs XX catalyst. \rg{The black dashed line represents $y=x$ i.e. when the catalyzed gap equals the uncatalyzed one.} We rescale each instance by the problem gap $\Delta_0$ for a fair comparison. (a) $\Delta_{c_1}$ denotes the gap when we use the $H_{cXX}$ catalyst. (b) $\Delta_{c_2}$ denotes the gap when we add the $H_{cXXX}$ , $\Delta$ denotes the gap for the $H_c=0$ scenario. (c) shows the improvement on addition of $H_{cXXX}$ in addition to $H_{cXX}$ compared to just using $H_{cXX}$ The red and green circled points are analyzed in more details in the main text.}
\label{fig:erdrenyicat}
\end{figure*}
We plot the energy gap, \( \Delta \), during the anneal for the problem described above without a catalyst ($H_c=0)$, and with two types of catalysts ($H_c=H_{cXX}$ and $H_c=H_{cXX}+H_{cXXX}$) in Fig.~\ref{fig:XXXbetter}(b). Note that $H_{cXXX}$ consists of couplings as shown in the second diagram of the top row of Fig.~\ref{fig:possible cases}(a),  an example of which is shown via green dotted lines in Fig.~\ref{fig:XXXbetter}(a), i.e., it connects all cases of three vertices with exactly two problem edges between them. An example of coupling of $H_{cXX}$ is shown via black dotted lines in Fig.~\ref{fig:XXXbetter}(a). The addition of $H_{cXXX}$ alongside $H_{cXX}$ to this configuration offers an exponential increase in the energy gap compared to the non-catalyzed case.
While the $H_{cXX}$ also offers an improvement, a sharp dip in the energy gap indicates that it was not able to convert the transition to a crossover.

\paragraph{\textbf {\rg{Statistics of improvement in Erdős-Rényi graphs:}}} To show that this phenomenon is statistically true for random MWIS problems, we conclude this section by showing minimum energy gap data obtained by setting up the MWIS problem on an Erdős-Rényi graph ~\cite{nutricati2024enhancingenergygaprandom}. We generate \(10^4\) random instances of the problem, choosing the probability \(p = 0.5\) for an edge \(E\) to exist between two vertices \(V\) of the graph. The antiferromagnetic couplings \(J_{ij}\) are drawn randomly from a uniform distribution over \([1, 2]\) \rg{and are chosen to be different on different edges to simulate distinct spin-glass like problems}. The on-site weights \(w_j\) are selected randomly from a uniform distribution over \([0, 1]\). These selections ensure the MWIS criteria \(J_{ij} > \min(w_i, w_j)\) are satisfied.

We compute the minimum energy gap during the anneal protocol, denoted as \(\Delta\) for just this section to avoid cluttering, without the addition of any catalyst (\(H_c = 0\)). We also calculate the minimum energy gap \(\Delta_{c1}\) upon adding a catalyst, which includes an \(XX\) coupling on all the edges of the problem (\(H_{cXX}\)), and a second case \(\Delta_{c2}\), where \(XXX\) couplings are added on all groups of three vertices with only two edges between them (\(H_{cXXX}\)). This setup follows our guideline of not connecting frustrated loops with a single coupler.

Figure~\ref{fig:erdrenyicat} presents a scatter plot comparing these scenarios, with each instance scaled by the problem gap \(\Delta_0\) to ensure a fair comparison. From Figures~\ref{fig:erdrenyicat}(a) and (b), \rg{the first} key observation is that there is a high density of points above the dashed line, indicating an improvement in the energy gap due to the addition of catalysts. Furthermore, in Figure~\ref{fig:erdrenyicat}(b), the data points appear to be more significantly displaced above the dashed line compared to Figure~\ref{fig:erdrenyicat}(a), clearly demonstrating greater improvement when the $H_{cXXX}$ catalyst is added.

To further emphasize this point, Figure~\ref{fig:erdrenyicat}(c) plots the energy gap using \(H_{cXXX}\) versus the energy gap using \(H_{cXX}\). The clustering of points above the line with slope \(=1\) indicates that statistically, adding \(H_{cXXX}\) in addition to \(H_{cXX}\) will increase the energy gap. This analysis completes the previous, where we show a scenario in which using only \(H_{cXX}\) is insufficient to eliminate the phase transition, while using \(H_{cXXX}\) succeeds.

\rg{While the majority of instances show improvement in the energy gap upon introducing the catalyst term \( H_{cXX} \) (and $H_{cXXX})$, some points lie below the dashed line in Fig.~\ref{fig:erdrenyicat}(a) and (b). In a few cases, the gap even decreases by an order of magnitude. These reflect a genuine physical effect: the catalyst induces strong hybridization of excited states, which can unintentionally trigger a  possible first-order phase transition.

To illustrate this, consider the instance marked by the red circle in Fig.~\ref{fig:erdrenyicat} (details of the configuration in Appendix~\ref{app:appC}, Tables~\ref{tab:redcoup} and ~\ref{tab:redpot}). Here, the ground ($\ket{g}$) and first excited ($\ket{e_1}$) states of the problem Hamiltonian \( H_P \) differ by a Hamming distance of $7$. Without any catalyst, the minimum gap during the anneal is the problem gap, with no first-order transition occurring during the anneal. However, the \( H_{cXX} \) term couples states differing by Hamming distance 2: \( \langle h | H_{cXX} | h' \rangle \neq 0 \) only if \( |h - h'| = 2 \) where $h$ is the Hamming weight of the state. Henceforth, for the states discussed below, we shall label them via the eigenstate of $H_p$ they have the largest overlap with, as that will determine the dominant contribution to the hybridization since we are looking at perturbative crossings.

The catalyst does not directly couple the ground, $\ket{g}$ and first excited state, $\ket{e_1}$. However, the $\ket{e_1}$ and the second excited state ($\ket{e_2}$) differ by Hamming distance 2, allowing strong hybridization via \( H_{cXX} \). As a result, $\ket{e_1}$ hybridizes with $\ket{e_2}$ and creates a new state whose energy is significantly lowered. Meanwhile, the third excited state ($\ket{e_3}$) is the closest energy state with Hamming distance 2 of $\ket{g}$, but it lies farther in energy, resulting in weaker hybridization. The net effect is that the hybridized $\ket{e_1}$ and $\ket{e_2}$ state becomes lower in energy than the hybridized $\ket{g}$ and $\ket{e_3}$ state, and it becomes the instantaneous ground state of $H(s=s_c - \varepsilon), \quad \varepsilon \ll 1
$. However since the ground state of $H_p$ is $\ket{g}$, this leads to an avoided crossing and a first-order transition at $s_c$ in the presence of $H_{cXX}$. This is absent when \( H_{cXX} \) is removed or replaced with \( H_{cXXX} \), as we demonstrate in Appendix~\ref{app:badcat}, corroborating our understanding.

This mechanism is essentially the inverse of what we have discussed briefly for the toy model, where specifically curated $XX$-couplings can drastically improve the gap closing by lowering the ground state energy via coupling to nearby low-energy excited states, while the first excited state remains largely unaffected. In contrast, in these detrimental cases, the excited state couples strongly to nearby states and gets lowered more than the ground state. If the energy gets lowered such that it becomes smaller than the state with maximal weight on $\ket{g}$ (the ground state of $H_p$) then there will necessarily be an anti-crossing during the quantum annealing process. If the energy is lowered but not enough for the switch, there will still be an $O(1)$ energy gap but of lower magnitude. Furthermore, only if $\ket{g}$ and $\ket{e_1}$ differ by a large Hamming distance, $O(L)$, compared to the $n-$locality of $H_D$ and $H_c$, as in the example described before, the energy gap becomes $O(e^{-L})$ by the same mechanism described in Sec.~\ref{sec:expgap}. As evident from Fig.~\ref{fig:erdrenyicat}, such highly detrimental cases are statistically rare, since they require the excited state to strongly hybridize using $H_c$, more so than the ground state, and also lie at a large Hamming distance from the ground state.

The complementary example of drastic \textit {improvement} exists for this random graph (green point in Fig.~\ref{fig:erdrenyicat}, the parameters are detailed in Appendix~\ref{app:appC}, Tables~\ref{tab:greencoup} and ~\ref{tab:greenpot}), just like for the toy model. In this beneficial case the catalyst strongly hybridizes the ground state with the third excited state (at Hamming distance 2), while the first excited state remains relatively unaffected, resulting in gap enhancement. Since the catalyst in such a scenario is not required to directly connect the ground and first excited state which is separated by a Hamming distance of $O(L)$ , there is no direct connection between the size of the problem $L$ and the efficacy of such a catalytic improvement. 

Rather, the catalyst's effect depends sensitively on the proximity and Hamming distances of nearby energy levels, which depends on the distribution of $J_{ij}$ and $h_i$. Favorable distributions can lead to significant gap improvements even for large $L$.  \rgg{When the unperturbed gaps are already small, even modest perturbations from $H_c$ can produce significant changes through higher-order processes in perturbation theory. 
By contrast, a reduction of a large gap requires a much more specific coincidence: an excited state above the first excited level must be shifted downward with remarkable accuracy so that it becomes nearly resonant with the ground state. 
Such near-degeneracies are statistically rare in the regime of perturbative crossings. 

To formalize this, we employ a perturbative formalism (Schrieffer–Wolff/ Van-vleck). 
Let $H=H_0+H_c$, where $H_0$ is the un-catalyzed Hamiltonian, and project onto the subspace containing the two states which we assume to be involved in the transition $\{\ket{1}, \ket{0}\}$. 
The effective Hamiltonian governing the low-energy subspace is
\[
H_{\mathrm{eff}}(E)=
\begin{pmatrix}
E_0^{(0)}+\Sigma_{00}(E) & V_{01}^{{\rm eff}}=V_{01}+\Lambda_{01}(E)\\[4pt]
V_{10}^{{\rm eff}}=V_{10}+\Lambda_{10}(E) & E_1^{(0)}+\Sigma_{11}(E)
\end{pmatrix},
\]
where $E_0^{(0)}$ and $E_1^{(0)}$ are the un-catalyzed energy of the states, and the self-energy and induced couplings till second order of perturbation theory due to hybridization with other states $\{\ket{u}\}_{u\ge 2}$ are,
\[
\Sigma_{aa}(E)=\sum_{u\ge 2}\frac{|V_{au}|^2}{E-E_u^{(0)}}, \qquad
\Lambda_{01}(E)=\sum_{u\ge 2}\frac{V_{0u}V_{u1}}{E-E_u^{(0)}}.
\]
A detrimental reduction of the minimum gap requires a near-resonance of one of the states $\ket{u}$ such that
\[
|\Delta_{01}+\delta_{01}| \lesssim |V_{01}+\Lambda_{01}(E_1')|,
\]
where $\Delta_{01}=E_1^{(0)}-E_0^{(0)}$ and $\delta_{01}=\Sigma_{11}(E)-\Sigma_{00}(E)$. 
In the localized or perturbative regime, the effective couplings scale as 
\[
|\Lambda_{0u}(E)|\sim (1-s)^{h(u)/n}\ll 1,
\]
where $h(u)$ is the Hamming distance between the configurations labeled $0$ and $u$, 
and $n$ denotes the locality of the catalyst. 
Consequently, because we have assumed that $|\Delta_{01}|$ is much larger than $|V_{01}^{\mathrm{eff}}|$ (i.e. the un-catalyzed instance does not have a gap closure), such near-resonances and the associated substantial reductions of the minimum gap are exceedingly unlikely. Achieving a resonance would require satisfying the condition $\delta_{01} \approx -\Delta_{01}$ to a very high degree of precision.
By contrast, when $\Delta_{01}$ is already small, the probability of a perturbative correction significantly modifying the gap is much higher. 
This qualitatively explains why, in the random-graph examples, gap enhancements are statistically more frequent than detrimental cases.} Nevertheless, the specific catalyst structure which is strongly beneficial cannot be known a priori without solving the spectrum. Hence, we advocate a heuristic strategy: introduce \( n \)-local catalyst terms incrementally and monitor their effect on performance, stopping once the benefit saturates or reverses. This adaptive approach balances gain in performance with computational overhead.
}

\begin{figure*}
    \centering
\includegraphics[width=0.66 \columnwidth]{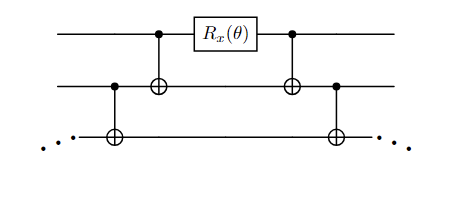}
\includegraphics[width=0.66\columnwidth]{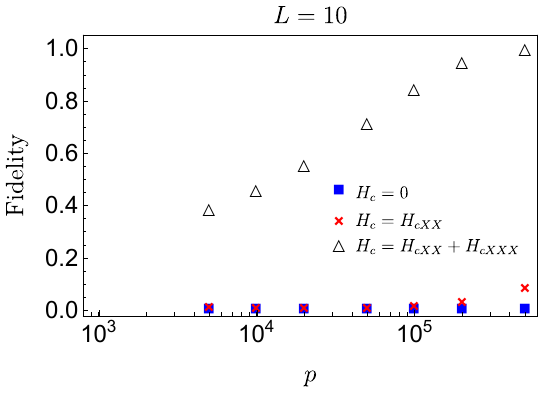}
        \includegraphics[width=0.66 \columnwidth]{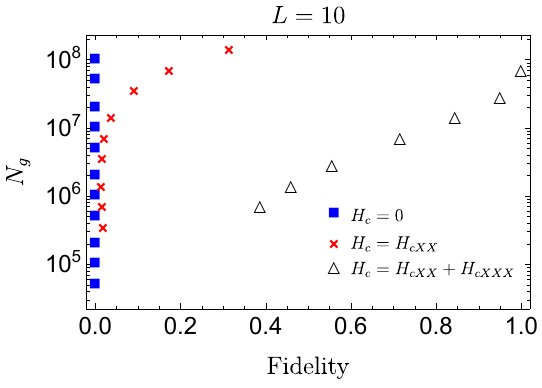}
    \caption{Left: Circuit representation of $XXX$ couplings. Middle: fidelity of Trotter-evolved wavefunction with exact ground state for different cases, for different time steps $\rgg{p}$. Right: number of gates $N_g$ required to obtain good fidelity with the exact wavefunction in different scenarios. Each Trotter step duration is $\delta t=0.05$.}
    \label{fig:XXXgate effect}
\end{figure*}
\subsection{\rgg{Circuit depth reduction in Gate-based  implementations}}
In this section, we discuss the integration of long-range couplings within a gate-based quantum annealer framework. A gate-based quantum annealer discretizes the continuous annealing protocol into a series of Trotter steps, allowing for stepwise simulation of the quantum evolution. The left panel of Fig.~\ref{fig:XXXgate effect} illustrates a quantum gate configuration that realizes the $XXX$ coupling. This is achieved by placing a two CNOT gates on either side of a Pauli-X rotation, denoted by the $R_x(\theta)$ gate. By stacking these CNOT gates in a “ladder” structure around the rotation, we effectively create multi-qubit couplings; the depth of this CNOT ladder determines the $n$ of the $n-$local catalysts.
The generalization to higher-order $XXX\hdots$ couplings by adding more rungs in the CNOT ladder is indicated by dots in the figure. An important advantage of this approach is that only nearest-neighbor CNOT couplings are needed, making the construction feasible even with hardware constraints on long-range connections.  

Further note that the gate count increases as $O(n)$ for a $n-$local coupling~\footnote{There are proposals for constant depth implementation as well, for example see Ref~72}. In fact, the number of gates required for one $n-$local coupling can be exactly computed as $1+2(n-1)$. Hence the total number of additional gates required to achieve \rgg{$p$} time steps to perform quantum annealing with a $n-$local catalyst is $N_g=\rgg{p}(v+ \sum_{n=2}^{n_{max}}[1+2(n-1)]g_n)$ where $v$ denotes the number of vertices of the problem and $g_n$ denotes the number of $n-$local gates of each type added. We shall now demonstrate, using the MWIS problem on the random graph defined in Fig.~\ref{fig:XXXbetter}(a), that for the same number of trotter steps and more importantly for the same total number of gates, adding $n-$local catalysts exponentially improve the fidelity of time evolved wavefunction with the exact ground state, compared to the uncatalyzed scenario. 

Discretizing the annealing protocol of Eq.~\eqref{eq:annealcat} into trotter steps with step duration $\delta t=0.05$, and starting from the ground state of the driver Hamiltonian as before, we simulate the fidelity of the time evolved wavefunction with the exact ground state of the problem classically using unitary evolution. In the middle panel of Fig.~\ref{fig:XXXgate effect}, we observe that for the case $H_c=0$, even after increasing the time steps to $\rgg{p}=10^6$ we see no improvement in the fidelity, \rgg{which remains effectively $0$. In contrast, while the fidelity using the $H_{cXX}$ catalyst stays near $0$ until about $\rgg{p}\le 2\times 10^5$, we see a clear upward trend around $p = 5 \times 10^5$.} Finally, for the $H_{cXX}+H_{cXXX}$ case we observe a clear upward trend already from  $O(10^4)$ Trotter steps and a near perfect fidelity $(99.9\%)$ is achieved around $\rgg{p}= 5 \times 10^5$ Trotter steps.

However, just comparing the behaviour with $\rgg{p}$ does not give the full picture as each time step involves $N_g/\rgg{p}$ gates ($1-$ and $2-$ qubit) which is different in each of the scenarios and is based on the $n$ of the $n-$local catalysts added. Thus, in the right panel of Fig.~\ref{fig:XXXgate effect} we provide the comparison of Fidelity with $N_g$,  the total number of gates involved in the quantum computation. This clearly reveals the superiority of the approach of adding $n-$local catalysts during the anneal protocol by demonstrating orders of magnitude reduction in the number of gates required to achieve different levels of fidelity without and with different catalysts  for the system described above. For example, to achieve $95\%$ Fidelity in this anneal, if we include $H_{cXXX}$ we can do so via $10^7$ gates, whereas just for $H_{cXX}$ we cannot achieve even $50\%$ fidelity with $10^8$ gates. The situation is completely hopeless if we do not add the  catalyst where the Fidelity of the wavefunction is practically nil for $10^8$ gates. To summarize, $n-$local catalysts can be created in polynomial complexity by $2-$local gates, and provide \rgg{order of magnitude} improvement in quantum annealing time and cost, thus reducing the need for resources. While we have shown just one example here, any of the situations (points in Fig.~\ref{fig:erdrenyicat}) can be similarly tested and will statistically show similar results.
This result establishes a powerful strategy for achieving the most accurate ground state using a fixed total number of gates. Initially, the ground state should be prepared without the catalyst. Subsequently, the gates can be systematically rearranged to incorporate the non-local couplings in the hierarchical sequence outlined in Fig.~\ref{fig:possible cases}. By iteratively refining the configuration until no further energy reduction is observed, this approach ensures an optimized ground state preparation. \rg{This systematic method maximizes the utility of available resources.}

\section{Discussion}

In the realm of adiabatic quantum annealing, first-order transitions are particularly challenging to mitigate due to the nature of the states involved. These states are usually separated by a large Hamming distance, implying that they are significantly different in terms of their configuration. However, direct-tunnel couplings are
the key to successfully open up a gap during a first-order phase transition. That is, the quantum fluctuations which directly connect the initial and final states across the transition, can most effectively create an energy gap. We illustrate the efficacy of such couplings as a catalyst showcasing a versatile implementation protocol. However, identifying the exact coupling a priori is typically impossible due to the complexity of the quantum system, unless the problem structure has a lot of symmetries. As a result, non-direct but additional quantum fluctuations become essential for approximating the necessary conditions for these direct-tunnel couplings.

These extra fluctuations \textit{must} be $n$-local, providing many-body quantum fluctuations stronger than those generated by the system's intrinsic driver. Among them, we show that there exists effective couplings with $n$ much smaller than the Hamming distance between the two states across the transition, but they are rare and difficult to identify due to the super-exponential number of possible configurations. 
However, we observe that the effectiveness of such catalysts \rg{does} not scale with problem size \rg{for the toy model, and possibly even for random graphs as suggested by small $L$ numerical simulations in Ref.~\cite{nutricati2024enhancingenergygaprandom} and our heuristic arguments based on the statistical analysis of problem instances on  Erdős-Rényi graphs}, opening up significant opportunities to design practical methods that leverage this property. 

This mirrors an approach used in classical simulations, where Monte Carlo algorithms addressing critical slowing down—such as the Swendsen-Wang~\cite{PhysRevLett.58.86} and Wolff cluster algorithms~\cite{PhysRevLett.62.361}—employ cluster flips to mimic the system's natural correlations near critical points. This strategy provides the resonant perturbations needed to overcome energy barriers. Our study highlights the potential for a quantum analogue by analyzing all configurations of catalyst couplings in small $L$ toy systems with inherent structural symmetry. We find that a limited number of small $n$-qubit catalytic couplings achieve exponential improvements for smaller systems and, remarkably, remain effective at mitigating phase transitions as system size increases, demonstrating robustness and efficiency. 

Using the toy models as microcosms of random graphs, we show that the $n$-local catalysts create an additional effective bridge between states separated by a large Hamming distance across the transition in MWIS problems, something not inherently provided by the standard driver Hamiltonian. As such, these catalysts are expected to significantly accelerate the annealing process even in larger graphs than those considered here. Verifying this effect in larger and more complex random graphs could offer valuable insights, representing an exciting direction for future research.

\rg{We have further demonstrated that product catalysts, which directly connect two states separated by the maximal Hamming distance in a problem, serve as examples of direct-coupling catalysts in the toy bipartite model. These catalysts non-perturbatively couple the ground and first excited state across the transition and are agnostic to whether they are stoquastic or non-stoquastic.} However, $n-$local catalysts are sensitive to this distinction. Stoquastic catalysts create a more systematic improvement by selectively lowering the energy of the instantaneous ground state involved in the transition, and if they minimally impact (or adversely impact) the higher energy level, a gap opens up. Non-stoquastic catalysts, on the other hand, introduce oscillating perturbations, resulting in highly system-dependent behavior.

While it is challenging to pinpoint the exact structures of direct-coupling catalysts, it is possible to identify structures that are ineffective for solving MWIS problems. Using them, we developed a hierarchical framework of catalyst connections tailored for MWIS problems, potentially extending this methodology to other types of problems as well. \rgg{In practice, one may progressively introduce $n$-local catalyst terms, 
starting from low locality and increasing $n$ until either the hardware or algorithmic limits are reached, 
or no further reduction in the final energy is observed. 
This hierarchical approach provides a systematic way to identify effective catalysts without full optimization. 
We emphasize, however, that increasing $n$-locality does not guarantee a monotonic improvement of the energy gap: 
beyond a certain point, additional multi-body terms can make the system more complex and may even reduce performance, 
as illustrated in some of the random-graph examples in the main text, where an $XX$ catalyst induces an additional phase transition. 
However, as discussed there, such detrimental cases are expected to be statistically rare. 
The proposed prescription is therefore meant as a practical and adaptive heuristic, 
not as a claim of universally monotonic enhancement.}

We further discuss gate based quantum circuit implementation of our proposed anneal protocol, where we show that, crucially, the $n-$local couplings can be generated by a circuit depth $O(n)$. Using this we demonstrate that the exponential reduction of anneal time translates into an order of magnitude reduction in number of gates necessary to obtain high fidelity solutions, offering a leap in efficient quantum computing. $n-$local catalysts induce entanglement across different sites in the problem graph, providing an additional source of entanglement alongside that generated by problem interactions. This entanglement is crucial for preventing gap closures. 

\rg{Our perturbative framework is not restricted to the specific toy model discussed here. In principle, it should apply to any system where perturbative crossings occur between low-lying energy states during the anneal, including exact cover~\cite{atshulerroland}, k-SAT~\cite{Farhi2011} and other spin glass like models~\cite{PhysRevLett.105.167204}. This is worth investigating in detail in a future work. However, what will differ is the procedure for reducing catalyst couplings at certain $n-$localities described in this work which specifically applies to MWIS-type problems. As a result, the polynomial reduction in the number of catalyst terms may not hold or may need to be replaced by a different rule in other problem classes. Nevertheless, we anticipate that the effectiveness of stoquastic catalysts should carry over to these problems as well.}

\rg{There are alternative catalyst terms which could be used in $H_c$, for example, involving $\sigma^y$ operators, but the situation can get more nuanced. Recall that $H_D$ is still $\sum_i\sigma^x_i$ and thus these terms render the Hamiltonian complex-valued, and therefore the Perron–Frobenius theorem no longer applies. Consequently, the sign structure of the perturbative series becomes indefinite, akin to the behavior of non-stoquastic catalysts, making analytic predictions based on low-order perturbation theory more challenging. However, $\sigma^y$ catalysts, just like $\sigma^x$ catalysts, can still connect computational basis states differing by large Hamming distances and may provide similar gap-opening benefits. However, in our numerical experiments on the toy model, we found $\sigma^x$ catalysts to outperform ones with $\sigma^y$ terms. Some work also has been done to study the effect of $\sigma^z$ catalysts~\cite{hattori2025controlled}, but the mechanism of energy gap improvement there is different, as it does not connect states but rather changes the diagonal terms in the $H(s)$. A thorough analysis of which setup is most efficient is beyond the scope of the present work.}

While some research has explored the utility of multipartite entanglement in quantum annealing~\cite{Batle2016}, a general understanding remains elusive, and we intend to investigate the geometry of additional quantum correlations that prove effective in future work. We also note that in approaches departing from strict adiabaticity, such as the Quantum Approximate Optimization Algorithm (QAOA) and Quantum Random Walks (QRW), gates are applied for extended durations and subsequently optimized, thereby inducing entanglement across different regions of the problem. \rg{Such optimization may yield additional speedups in quantum state preparation beyond what is demonstrated here, especially as recent results suggest that optimal control~\cite{Pecci_2024} is particularly effective.} The $n$-local catalyst provides an adiabatic pathway to generate this entanglement, albeit at the cost of multi-qubit control. \rg{The observed gate count reduction from $n$-local catalysts can also complement QAOA-like techniques. Since we find that introducing these new classes of gates significantly enhances state preparation efficiency even without parameter tuning, we expect additional speedups may be achieved through further optimization. While we do not claim that the addition of $n$-local gates without further optimizations can already outperform classical algorithms or the best QAOA protocols, our results provide a new lens through which we can understand and engineer quantum speedups. We believe this insight will serve as a foundation for enhancing existing quantum state preparation strategies and developing more powerful hybrid quantum algorithms.}
\begin{acknowledgements}
    RG, LN, PAW and SB thank EPSRC grant EP/Y004590/1 MACON-QC for support.
\end{acknowledgements}

\bibliography{ref}

\appendix
\rg{\section{Derivation of the MWIS Spin Glass Hamiltonian}
\label{app:MWISising}

We provide here a step-by-step derivation of the spin-glass Hamiltonian corresponding to the MWIS problem on a graph \( G = (V, E) \) with node weights \( w_i \).

The MWIS problem can be cast as a binary optimization problem over variables \( x_i \in \{0,1\} \), where \( x_i = 1 \) indicates that vertex \( i \) is included in the independent set. The optimization function with a soft constraint for independence is:
\begin{equation}
\mathcal{H}(x) = \sum_{i \in V} w_i x_i - \sum_{(i,j)\in E} J_{ij} x_i x_j,
\label{eq:MWIS_binary}
\end{equation}
where \( J_{ij} > \max(w_i, w_j) \) are penalty coefficients that ensure neighboring vertices are not both included in the independent set. The task is to obtain the configuration of $\{x_i\}$ which maximizes this function.

To map this to an Ising spin-glass Hamiltonian suitable for quantum annealing, one introduces spin variables \( s_i \in \{-1, 1\} \) via the transformation:
\begin{equation}
x_i = \frac{1 + s_i}{2}, \quad \text{so that} \quad x_i x_j = \frac{1 + s_i + s_j + s_i s_j}{4}.
\end{equation}

Substituting into Eq.~\eqref{eq:MWIS_binary}, we compute the two components of the Hamiltonian.

\paragraph{Objective Term:}
\begin{align}
\sum_{i \in V} w_i x_i &= \sum_{i \in V} w_i \cdot \frac{1 + s_i}{2} \\
&= \frac{1}{2} \sum_{i \in V} w_i + \frac{1}{2} \sum_{i \in V} w_i s_i.
\end{align}

\paragraph{Penalty Term:}
\begin{align}
-\sum_{(i,j)\in E} J_{ij} x_i x_j &= -\sum_{(i,j)\in E} J_{ij} \cdot \frac{1 + s_i + s_j + s_i s_j}{4} \\
&= -\sum_{(i,j)\in E} \left( \frac{J_{ij}}{4} + \frac{J_{ij}}{4}s_i + \frac{J_{ij}}{4}s_j + \frac{J_{ij}}{4}s_i s_j \right).
\end{align}

Collecting the terms we obtain,
\\

Constant Term:
\[
\frac{1}{2} \sum_{i \in V} w_i - \sum_{(i,j)\in E} \frac{J_{ij}}{4},
\]
Linear Term (effective field \( h_i \)):
\begin{equation}
h_i = \frac{1}{2} w_i - \frac{1}{4} \sum_{j \in \rm{nbr}(i)} J_{ij}
\end{equation}
and Quadratic Term (coupling \( J_{ij} \)):
\[
-J_{ij}^{\text{(Ising)}} = -\frac{1}{4} J_{ij} \quad \text{for all } (i,j) \in E.
\] The crucial thing to note here is to avoid the double counting of all the edges in the graph.

\paragraph*{Final Spin-Glass Hamiltonian:}
Ignoring the constant shift in energy and multiplying the overall Hamiltonian by $4$, the final Ising Hamiltonian becomes:
\begin{eqnarray}
H &=&4 \sum_{i \in V} h_i s_i - 4\sum_{(i,j)\in E} J_{ij}^{\text{(Ising)}} s_i s_j \label{eq:eqinterising} \\
&=& \sum_{i \in V} \left( 2 w_i -  \sum_{j \in \rm{nbr}(i)} J_{ij} \right) s_i - \sum_{(i,j)\in E} J_{ij} s_i s_j. \nonumber
\end{eqnarray}
Recall that we are looking to find the configuration which maximizes Eq.~\eqref{eq:eqinterising}, which means it is the same configuration which minimizes $-H$. This is the ground state eigenfunction of the Ising Hamiltonian in Eq.~\ref{eq:isingMWIS} of the main text. For our simulations we generate random weights and couplings which satisfy  \( J_{ij} > \max(w_i, w_j) \) for each graph instance. While the solution of the MWIS problem itself only depends on the graph structure and not on the specific values of the couplings $J_{ij}$, we have randomized them as well. Thus in essence we generate spin-glass-like problems directly on the random graph but obeying the MWIS constraint.}

\section{\rgg{Condition to ensure first order phase transition in the bipartite model}}
\label{sec:appA}
Recall the MWIS Hamiltonian in Eq.~\eqref{eq:isingMWIS}, 
\begin{equation}
    H_p = \sum_{ij \in E} J_{ij} \sigma^z_i \sigma^z_j + \sum_{i \in V} \left(\sum_{j \in \mathrm{nbr}_i} J_{ij} - 2 w_i \right) \sigma^z_i,
\end{equation}
for the toy model of Fig.~\ref{fig:toy}.
 We order the computational basis states such that the first four bits represent the spins in system $A$. \rgg{If we choose} $W_2>W_1$ in the toy model, the ground state is $\ket{\downarrow\downarrow\downarrow\downarrow\uparrow\uparrow\uparrow}$. \rgg{The Maximum independent set here is the set of \rg{three} up spins in system $B$.} Our goal is to find a condition when the first excited state of the problem Hamiltonian is a large Hamming distance away from the ground state, for reasons that will be clear at the end of this appendix. 

To find the first excited state, one can flip a spin in the left or right block, or flip more than one spin.
\begin{enumerate}
    \item \textbf{Flip spin 1}$ \ket{\downarrow\downarrow\downarrow\downarrow\uparrow\uparrow\uparrow} \rightarrow  \ket{\uparrow\downarrow\downarrow\downarrow\uparrow\uparrow\uparrow}$: 
    We consider $J_{ij}=J$ for simplicity. The energy cost for this spin-flip is 
    \begin{equation}
   (3 J + (3 J-2 w_1))-(-3 J-(3 J- 2 w_1)) =12 J-4 w_1=12J-  W_1.
    \end{equation}
    \item \textbf{Flip spin 5}$ \ket{\downarrow\downarrow\downarrow\downarrow\uparrow\uparrow\uparrow}  \rightarrow  \ket{\downarrow\downarrow\downarrow\downarrow\downarrow\uparrow\uparrow} $: 
    The energy cost for this spin-flip is 
    \begin{eqnarray}
    &&(4 J -(4 J-2 w_5))-(-4 J+(4 J- 2 w_5)) \nonumber \\
    &&=4 w_5=4W_2/3
    \end{eqnarray} 
     \item \textbf{Flip all spins}$ \ket{\downarrow\downarrow\downarrow\downarrow\uparrow\uparrow\uparrow}  \rightarrow   \ket{\uparrow\uparrow\uparrow\uparrow\downarrow\downarrow\downarrow} $: 
    Here, we can ignore the antiferromagnetic interactions which remain the same for both cases. Thus, we have,\\
    before:$-[(3J -2 w_1+3J -2 w_2+3J-2 w_3+3J-2 w_4)-(4J-2w_5+4J-w_6+4J-w_7)]=2(w_1+w_2+w_3+w_4-w_5-w_6-w_7)$\\
    after:$-2(w_1+w_2+w_3+w_4-w_5-w_6-w_7)$.\\
    The difference and thus the energy cost is $4(-w_1-w_2-w_3-w_4+w_5+w_6+w_7)$
    \begin{equation}
    =4(-\sum_{i=1\hdots4}w_i+\sum_{i=5 \hdots 7} w_i)=4 (W_2-W_1)
    \end{equation}
Finally we consider another case when we flip two spins,
\item \textbf{Flip spin $1$ and $5$}$ \ket{\downarrow\downarrow\downarrow\downarrow\uparrow\uparrow\uparrow} \rightarrow  \ket{\uparrow\downarrow\downarrow\downarrow\downarrow\uparrow\uparrow}$:
The energy cost for performing these flips is $J+(3J-2 w_1)+2J-(4J-2 w_5)-(-3J-(3J-2w_1)-4J+(4J-2w_5)$
\begin{equation}
=8 J -4 w_1+4 w_5
\end{equation}
Since $J > w_i$ by problem statement, \rgg{case 4} always has greater energy cost \rg{than case 2}.
\end{enumerate}
Since the energy cost of flipping more than one spin in a single partition is additive, we do not need to consider the other scenarios of flipping more spins in any of the subsystems. \rg{Additionally, as $J>w_1,w_5$, \rgg{it is clear that} \textbf{Case 2} and \textbf{Case 3} are the cases which can have comparable energies.} Finally, if \begin{equation}
(W_2-W_1) < {\rm min} (w_5,w_6,w_7)=W_2/3
\label{eq:condition}
\end{equation}
then the state which is farthest away from the ground state in Hamming distance becomes the first excited state of the toy problem.

\rgg{Upon introducing quantum fluctuations via the transverse-field driver 
\( H_D = \sum_i \sigma_i^x \) (for \( s \lesssim 1 \) in \( H(s) \)), 
the ground and first excited states of \( H_p \) can hybridize with other eigenstates, either directly or through higher-order perturbative processes. 
As a result, at some value \( s < s_c < 1 \), the ground state of \( H(s) \) may switch from having maximal overlap with the ground state of \( H_p \) 
to having maximal overlap with the first excited state of \( H_p \). 
This phenomenon is referred to as a \textit{perturbative crossing}.

Formally, let \( \{\ket{s_i}\} \) denote the computational basis states, which are eigenstates of \( H_p \) with energies \( E_i^{(0)} \). 
In the perturbative regime, the eigenstates of \(H(s)\) can be expanded as
\[
\ket{\psi_i(s)} \approx \ket{s_i} + \mathcal{O}(\|H_D\|),
\]
with perturbed energies
\[
E_i(s) \approx E_i^{(0)} + \delta E_i(s),
\]
where \( \delta E_i(s) \) arises from higher-order processes induced by \( H_D \).  
A perturbative crossing between the lowest two levels occurs when, for some \( s<s_c \),
\[
E_1^{(0)} < E_2^{(0)} 
\qquad\text{but}\qquad 
E_1(s) > E_2(s).
\]
Thus, as \( s \) is reduced from \(1\), the energetic ordering of the two states switches at a critical value \( s_c < 1 \), 
even though each \( \ket{\psi_i(s)} \) remains dominantly supported on its corresponding computational basis state \( \ket{s_i} \).

For the toy model, we consider a complete bipartite graph with partitions \(A\) and \(B\) of sizes \( (L+1)/2 \) and \( (L-1)/2 \), respectively.  
A perturbative crossing can be ensured by choosing the total weight on the smaller partition \(B\) to be slightly larger than that on \(A\), i.e.,
\[
W_B > W_A, 
\qquad 
|W_B - W_A| \ll 1,
\]
with
\[
W_A = \sum_{i\in A} w_i, 
\qquad 
W_B = \sum_{j\in B} w_j.
\]
This choice makes the two MWIS configurations nearly degenerate at \( s=1 \), 
so that higher-order perturbative corrections from \(H_D\) can reverse their ordering and produce a level crossing at some \( s_c<1 \).  
Qualitatively, this can be understood by noting that the transverse driver $H_D = -\sum_i \sigma_i^x$ perturbs the larger partition more strongly than the smaller one, since a greater number of spins contributes non-negligibly to the perturbation series.

Crucially, if an order parameter changes abruptly at \( s_c \), the system undergoes a first-order quantum phase transition.  
If the ground and first excited states of \( H_p \) differ by only a single spin flip, the transverse field directly mixes them and no such transition occurs.  
However, the chosen parameters ensure that the two states differ by the maximal Hamming distance \( L \).  
As discussed in Sec.~\ref{sec:expgap}, when the ground and first excited states switch roles under such circumstances, 
the minimum gap scales as \( O(e^{-L}) \), reflecting an exponentially small avoided crossing.

The term \textit{perturbative crossing} refers to the fact that the eigenstates near the transition can be described perturbatively around \( H_p \); 
however, the crossing itself is a non-perturbative, collective effect involving \(O(L)\) qubits.  
Thus, in this bipartite toy model, two elements jointly determine whether a perturbative crossing occurs:  
(1) the larger weight must lie on the smaller partition (necessary);  
(2) the classical energy difference must be sufficiently small (necessary). 
Once this difference satisfies the inequality in Eq.~\eqref{eq:condition}, 
the driver-induced corrections inevitably overturn the ordering of the two MWIS configurations.  
Hence, combined with (1), Eq.~\eqref{eq:condition} provides a sufficient criterion for such a crossing, 
consistent with the behavior shown in Fig.~\ref{fig:toy}.

The perturbative crossing also constitutes a first-order quantum phase transition, 
since the relevant order parameter changes discontinuously.  
To conclude, the first excited state behaves as a local minimum that the system can reach efficiently without a catalyst; 
however, transitioning from this local minimum to the true ground state requires tunneling through a non-perturbative barrier associated with a first-order transition.}

\begin{figure}
    \centering
    \includegraphics[width=0.8\linewidth]{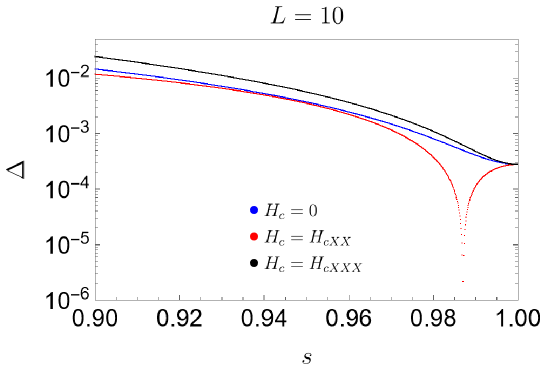}
    \caption{\rg{The variation of the energy gap $\Delta$ with rescaled time $s$ the instance labeled by the red circle in Fig.~\ref{fig:erdrenyicat}, showing the appearance of gap closure when $H_{cXX}$ is used.}}
    \label{fig:worsecatalyst}
\end{figure}

\rg{\section{Example where $H_{cXX}$ introduces a new transition}
\label{app:badcat}
In the main text we identified a case labeled by the red circle in Fig.~\ref{fig:erdrenyicat} where $H_{cXX}$ severely reduced the energy gap and performed an analysis of why this is the case. We claimed that the additional hybridization introduced by $2-$local $H_{cXX}$ is the culprit and the next higher $n-$local catalyst $H_{cXXX}$ does not show this effect. In Fig.~\ref{fig:worsecatalyst}, we show exactly this behaviour where we show that if $H_{cXX}$ is removed from the catalyst hierarchy, an application of just $H_{cXXX}$ does not cause a gap closing, corroborating our theory.}
\section{Non-stoquastic catalysts}
\label{app:appB}
\begin{figure}
    \centering
    \includegraphics[width=0.45\columnwidth]{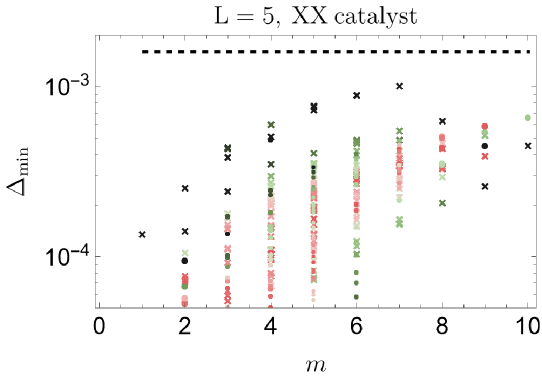}
    \includegraphics[width=0.45\columnwidth]{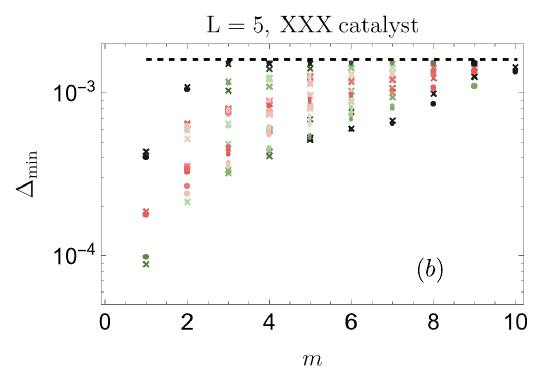}
    \includegraphics[width=0.45\columnwidth]{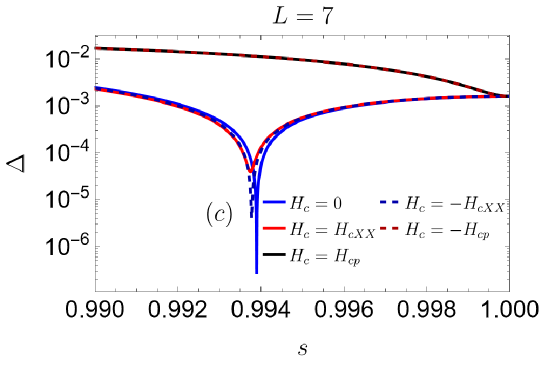}
    \includegraphics[width=0.45\columnwidth]{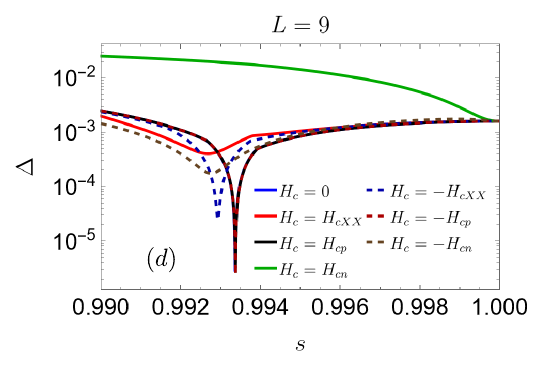}
    \caption{A comparison between non-stoquastic and stoquastic catalysts. (a) shows a comparison of the minimum energy gap of the bipartite toy model in Fig.~\ref{fig:toy}, $\Delta_{\rm min}$, when we add XX couplings with $-$ve sign (stoquastic), shown by $\times$, with addition of XX couplings with $+$ve sign (non-stoquastic) shown by filled circles. (b) shows the same data as (a) on addition of XXX couplings. (c) shows a comparison of how the energy gap for the same problem behaves during the anneal on addition of stoquastic vs nonstoquastic $H_{cXX}$ and $H_{cp}$. (d) shows a similar plot as (c) for the tripartite graph of Fig.~\ref{fig:tripartite} for the same choices of catalyst as in Fig.~\ref{fig:tripartite}(b) but with different signs. Note that the catalysts used in main text had $-$ve sign, i.e. were stoquastic. }
    \label{fig:nonstoq}
\end{figure}
In this appendix, we provide a brief comparison of the behavior of stoquastic versus non-stoquastic catalysts for the problems considered in the main text. Our main conclusion is that if one uses a direct-tunnel coupling with a product catalyst, the sign of the catalyst does not matter, \rg{whereas for other catalyst couplings}, even if it is a direct-tunnel one, stoquastic catalysts offer a much more systematic pathway to improve the energy gap, while a non-stoquastic catalyst introduces a problem-dependent improvement. In Fig.~\ref{fig:nonstoq}, we provide evidence supporting these claims. In Fig.~\ref{fig:nonstoq}(a), we observe that the number of exceptionally effective catalysts is much greater in the stoquastic case, denoted by $\times$, compared to the non-stoquastic case, denoted by $\bullet$. Otherwise, there is limited distinction between the two types of catalysts. This is further seen for $XXX$-catalysts, where due to the higher degree of connections relative to the system size, the lines are further blurred between the two types of catalysts.

To complete our analysis, we also consider the specific cases of $H_{cXX}$ and $H_{cp}$ for the bipartite graph in Fig.~\ref{fig:nonstoq}(c), where we show that the direct-tunnel coupling induced by $H_{cp}$, which connects states maximally apart in Hamming distance, remains unaffected by the sign, whereas the $H_{cXX}$ is adversely affected by introducing non-stoquasticity.  On  the other hand, for the tripartite case shown in Fig.~\ref{fig:nonstoq}(d), where now the direct-tunnel coupling is induced by $H_{cn}$ which still has perturbative effects on the energy states as it does not connect states maximal Hamming distance apart,  the stoquastic choice is preferable. This can again be  explained via the perturbative arguments of the main text. We therefore conclude that non-stoquasticity provides no additional improvement for the problems considered in this work.

\rg{\section{Mitigating the transition in the ferromagnetic $p$-spin model}
\label{app:p-spin}
The ferromagnetic $p$-spin model is defined as
\begin{equation}
    H_p = -L\left(\frac{1}{L} \sum_{i=1}^L \sigma^z_i\right)^p,
    \label{eq:p-spin}
\end{equation}
where $L$ is the system size. It is known~\cite{Jorg_2010,Nishimori2017} that to prepare the ground state of such a model using adiabatic quantum annealing with the standard transverse field driver $H_D = -\sum_{i=1}^L \sigma^x_i$, the system undergoes a phase transition at $s = s_c(p)$, where the ground state changes from paramagnetic to ferromagnetic. Furthermore, for $p \geq 3$, this transition is of first order with an exponentially closing gap~\cite{Jorg_2010}.

In the main text, we discussed the origin of this exponentially closing gap and clarified that this is not the type of gap closing which we primarily aimed to address in this work. Nevertheless, in this appendix, we explore the consequence of applying the catalyzed annealing procedure to this model and demonstrate how our understanding aligns with the well-known superiority of the non-stoquastic catalyst in opening up the gap in the $p$-spin model~\cite{Nishimori2017}.

In the presence of a catalyst, the annealing protocol used is
\begin{equation}
    H(s) = s H_p + (1 - s) H_D + s(1 - s) H_c,
    \label{eq:protocol}
\end{equation}
where $H_c = \pm H_c^1=\pm L\left( \frac{1}{L} \sum_{i=1}^L \sigma^x_i\right)^k$. Note that $H_c$ is all-to-all coupled, consistent with the all-to-all nature of $H_p$ and in line with the catalyst constructions considered in this work.

In Fig.~\ref{fig:p-spin data}, we show the behavior of the energy gap during the annealing process for the cases without a catalyst ($H_c = 0$), with a stoquastic catalyst ($H_c = -$), and with a non-stoquastic catalyst ($H_c = +$) for three different choices of $p$ and $k$ (we leave out writing $H_c^1$ for brevity). We first observe that for the no-catalyst case, the energy gap closes during the anneal in all three parameter settings considered. A more detailed investigation (bottom panels) confirms that this gap is exponentially small in system size $L$.

When a stoquastic catalyst is added ($H_c = -$), the energy gap does not reopen; rather, the gap minimum shifts toward larger $s$ values. Moreover, the system-size scaling becomes even more severely exponential with $L$. On the other hand, with a non-stoquastic catalyst ($H_c = +$), the gap closes more smoothly around the critical point in all scenarios. This smoothness increases progressively as we move through the parameter sets $p = 3, k = 3$, then $p = 5, k = 3$, and finally $p = 5, k = 5$. In the last case, the structure of the gap closing is reminiscent of power-law decay of the gap with $|s-s_c|$ near the critical point.

The bottom panels of Fig.~\ref{fig:p-spin data} further demonstrate the superior performance of the non-stoquastic catalyst: the gap scaling with $L$ is slower in all three cases, and for $p = 5, k = 5$, it transitions to a power-law scaling in $L$ (verified but not shown), consistent with the findings of Nishimori et al.~\cite{Nishimori2017}.

This power-law scaling and the smoother behavior of the energy gap around the critical point suggest that the transition may become second order under non-stoquastic catalysis. We do not analyze the transition order further in the present work as it has been done before~\cite{Nishimori2017}.

\begin{figure*}
    \centering
    \includegraphics[width=0.6\columnwidth]{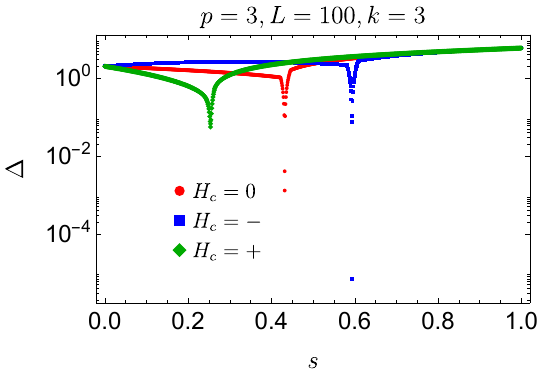}
    \includegraphics[width=0.6\columnwidth]{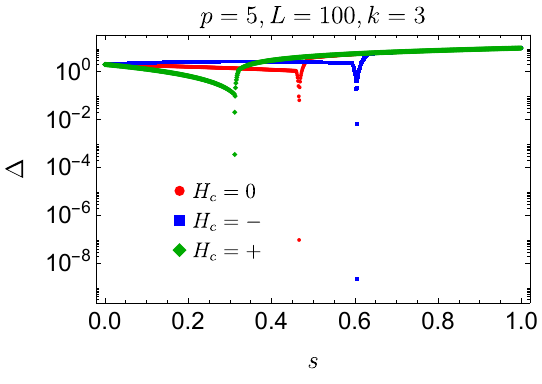}
    \includegraphics[width=0.6\columnwidth]{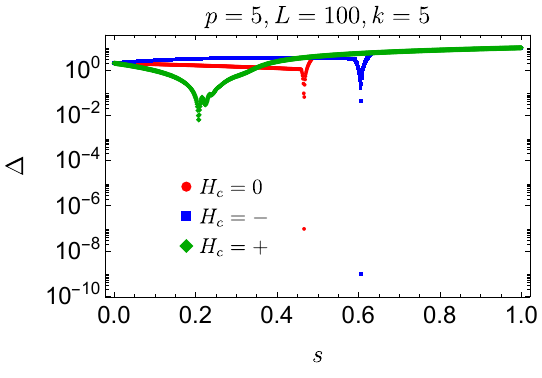}
    \includegraphics[width=0.6\columnwidth]{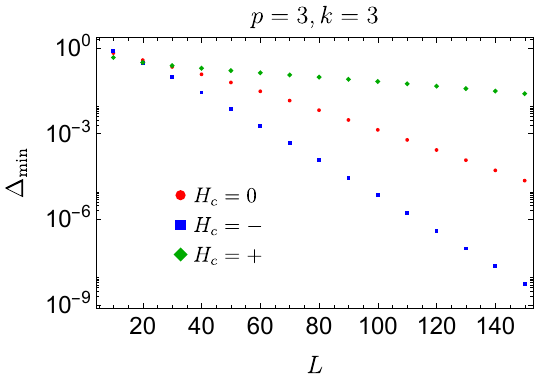}
    \includegraphics[width=0.6\columnwidth]{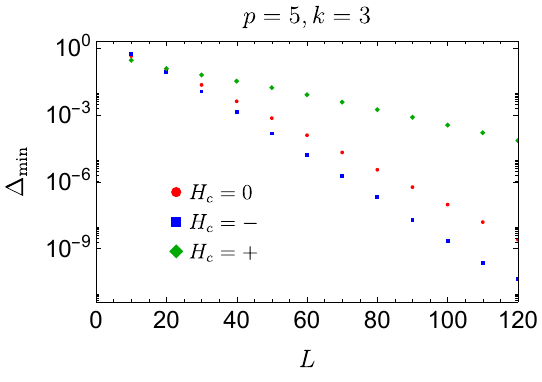}
    \includegraphics[width=0.6\columnwidth]{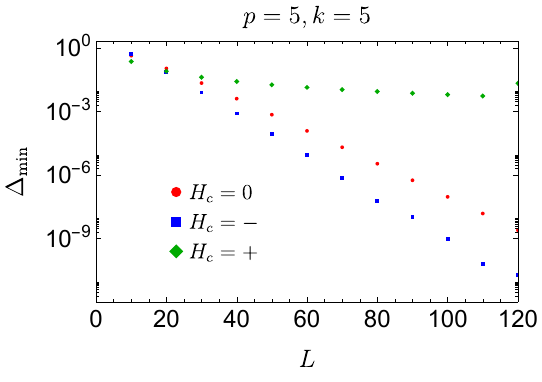}
    \caption{Data showing the effect of stoquastic ($H_c=-$) and non-stoquastic ($H_c=+$) catalysts on preparation of ground state of $p-$spin model using quantum annealing via Eq.~\eqref{eq:protocol}. Left panels: For $p=\rgg{k}=3$ in the p-spin model described in Eq.~\eqref{eq:p-spin}, in the top panel we show the instantaneous gap between the ground and first excited state during the anneal and in the bottom panel we show the scaling of the minimum energy gap $\Delta_{min}$ during the anneal with system size $L$. Middle and right panels show the same results for $p=5,k=3$ and $p=5,k=5$ respectively.}
    \label{fig:p-spin data}
\end{figure*}

\begin{figure*}
    \centering
    \includegraphics[width=0.45\columnwidth]{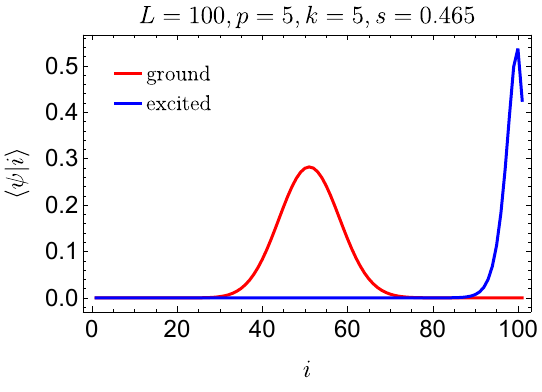}
    \includegraphics[width=0.45\columnwidth]{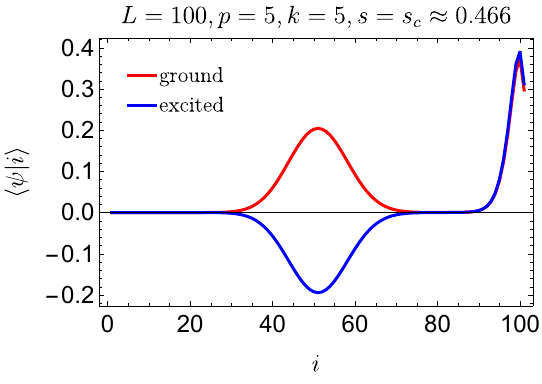}
    \includegraphics[width=0.45\columnwidth]{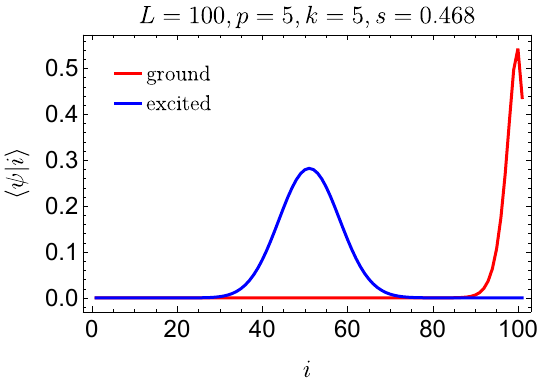}
    \includegraphics[width=0.45\columnwidth]{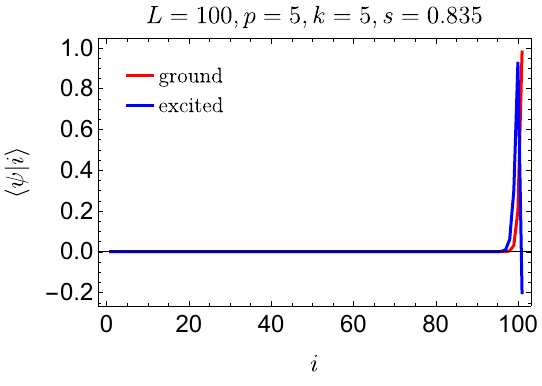}\\
    \includegraphics[width=0.45\columnwidth]{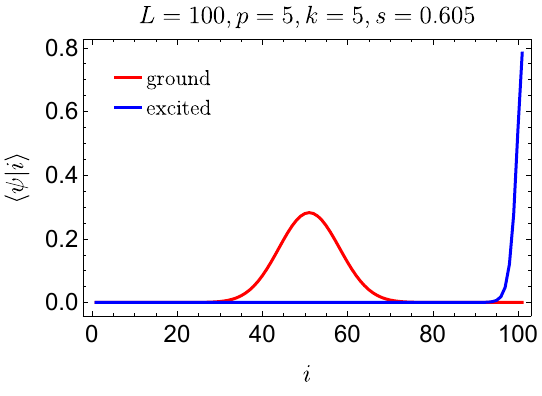}
    \includegraphics[width=0.45\columnwidth]{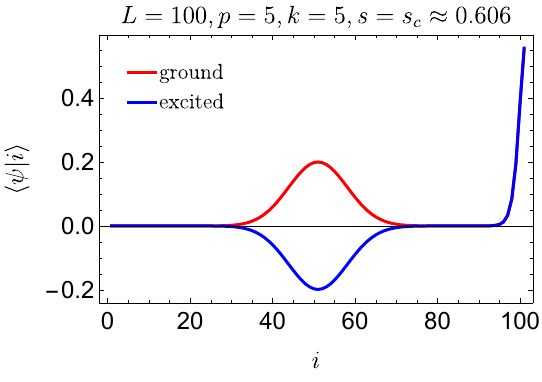}
    \includegraphics[width=0.45\columnwidth]{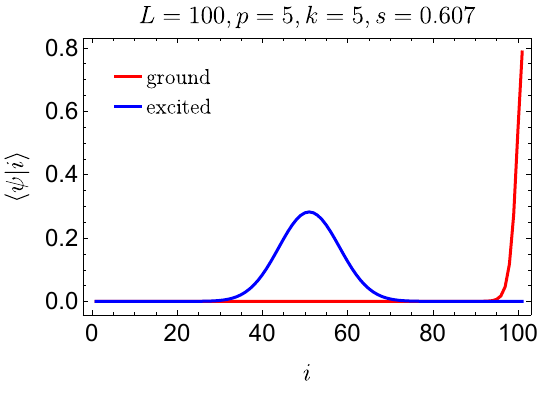}
    \includegraphics[width=0.45\columnwidth]{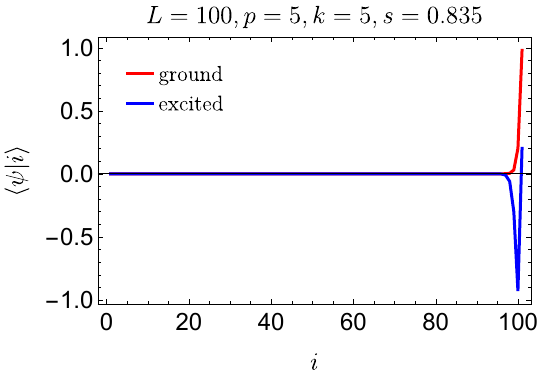}\\
    \includegraphics[width=0.45\columnwidth]{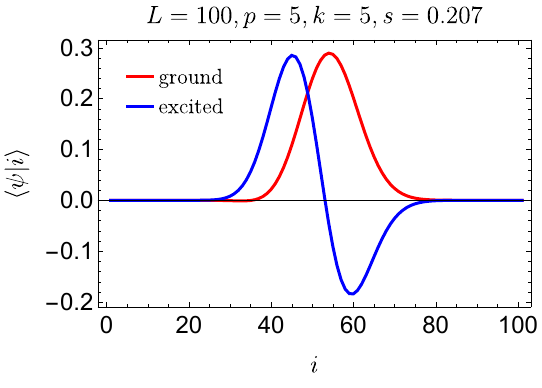}
    \includegraphics[width=0.45\columnwidth]{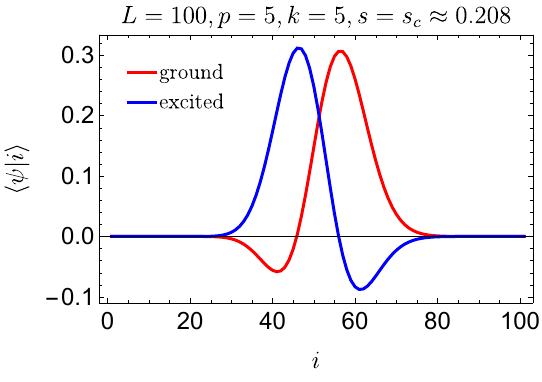}
    \includegraphics[width=0.45\columnwidth]{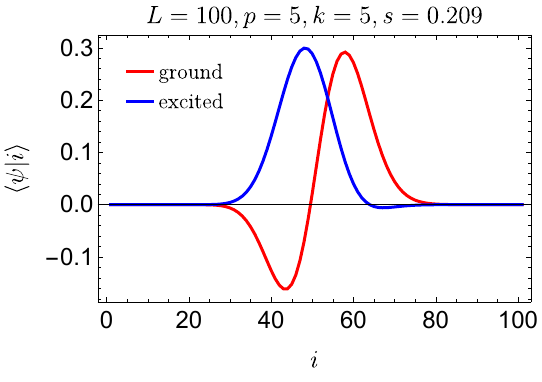}
    \includegraphics[width=0.45\columnwidth]{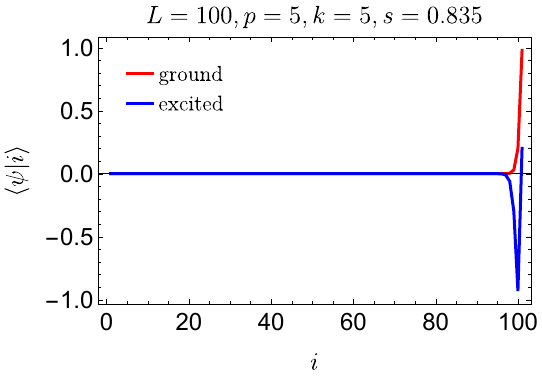}
    \caption{Distribution of eigenstate ($\ket{\psi}$ weights on basis states $\ket{i}$, for the basis defined in Eq.~\eqref{eq:Dicke states}. Top: For the no-catalyst case $H_c=0$, the critical point is at $s_c=0.46659312033$. The panels from left to right shows the behaviour of the ground and excited states just before, at, just after and away from criticality respectively. The middle panel does the same but for $H_c=-$, where $s_c=0.605792442974086$. The bottom panel shows for the non-stoquastic case for which $s_c=0.20794648993$. Clear differences in behaviour around criticality is seen for the ground and excited states. }
    \label{fig:eigenstate_data_p-spin}
\end{figure*}

However, we now explore why the sign of the catalyst has a qualitatively different effect on the phase transition in the $p$-spin model compared to the examples discussed in the main text. The key distinction lies in the nature of the transition itself: here, it is a paramagnetic-ferromagnetic transition—unlike the perturbative crossing phenomena addressed in the main text.

In this case, the mechanism by which the energy gap increases is not due to a direct connection between the ground and first excited states, as discussed in the main text. Instead, the non-stoquastic catalyst changes the energetics of the eigenstates near the transition, whereas the stoquastic catalyst does not. We illustrate this in Fig.~\ref{fig:eigenstate_data_p-spin} for the most dramatic case of $p=5$, $k=5$.

To compute the eigenstates, we use the fact that the total angular momentum operator $\mathcal{L}^2$ commutes with $H(s)$, rendering the Hamiltonian block-diagonal in the rotated basis. The ground and first excited states lie within the maximum angular momentum sector $\mathcal{L} = L/2$, whose basis states are the Dicke states. Exploiting this symmetry, we can efficiently diagonalize systems of size $L \sim \mathcal{O}(100)$~\cite{Jorg_2010}.

The Dicke states are defined as:
\begin{equation}
    \ket{i} = \ket{\mathcal{L} = L/2; \mathcal{L}_z = m_z},
    \label{eq:Dicke states}
\end{equation}
where $-L/2 \leq m_z \leq L/2$ and \rgg{$\ket{i}$ denotes a Dicke basis state}. The paramagnetic state corresponds to a Gaussian-weighted superposition centered around $i = L/2 + 1$, while the ferromagnetic state corresponds to $i = L + 1$.

Just before the critical point, in the region where $H_D$ dominates, all three scenarios yield a paramagnetic ground state. The difference appears in the nature of the first excited state: for $H_c = 0$ and $H_c = -$, the first excited state closely resembles the ferromagnetic ground state. However, for $H_c = +$, it is a different state having significant overlap with the paramagnetic ground state, even near criticality. At the transition point, we observe a superposition, followed by a level crossing between the ground and first excited states. For $s \gg s_c$, all cases show similar behavior, up to a global phase.

This distinction softens the transition in the presence of the non-stoquastic catalyst. The off-diagonal matrix elements of $H_D$ connect Dicke states $\ket{i}$ and $\ket{i \pm 1}$, since $\sum_i \sigma^x_i$ flips individual spins. The $k$-local catalyst $H_c$ allows connections up to $\ket{i \pm k}$. However, even with the catalyst, direct connection between the paramagnetic and ferromagnetic states remains exponentially suppressed due to negligible overlap between states such as $\ket{i = L/2 + 6}$ and $\ket{i \sim L}$ which need to be connected. The perturbative improvements possible are also limited, since the higher excited states have significant energy separation ($>O(1)$) with the ground and first excited states. This explains the persistence of a first-order transition in the stoquastic case.

In contrast, for $H_c = +$, the first excited state exhibits strong overlap with the ground state within a Hamming distance of $\sim 5$, enhancing the off-diagonal matrix element as per Eq.~\eqref{eq:matrix}, and thus increasing the energy gap. A detailed scaling analysis of the energy gap could be pursued, but lies beyond the scope of this work.

We can, however, explain the qualitative change in the first excited state due to different catalyst signs. At $s = 0$, the eigenstates are tensor product eigenstates of $\sigma^x$, such as $\ket{+++\dots}$,  $\ket{++-\dots}$ etc. expressed in the Dicke basis. When the catalyst is introduced without $H_p$, it modifies the energy of these states. For a $k$-local catalyst $H_c$ with the same sign as $H_D$, the effect is additive and lowers all energy levels. This energetically favors the paramagnetic state for larger $s$, pushing the para-ferro transition to larger $s$ and reducing the off-diagonal coupling in  basis of $\sigma^x$, as it scales with $(1 - s)$. As a result, the energy gap becomes even smaller.

However, when $H_c = +$, the catalyst has the opposite sign to the driver, introducing competition between ground and excited states. Simple algebra shows that at
\begin{equation}
    s_c = \frac{2 L^{k-1}}{(2 - L)^k - (-L)^k},
\end{equation}
the ground and first excited states exchange roles. Note that this derivation neglects $H_p$ and considers the classical part of the model where all terms commute. For $k = 5$ and $L = 100$, this gives $s_c \approx 0.208162$, remarkably close to the actual critical point $s_c \approx 0.207946$ when $H_p$ is included (see Fig.~\ref{fig:eigenstate_data_p-spin}). Numerical analysis (not shown) confirms that $H_p$ acts perturbatively in this limit.

In the Dicke basis, since both the ground and first excited states now have significant overlap, and $H_p$ is diagonal in this basis, a nonzero gap opens up. Once the ground state transitions away from the paramagnetic state, no further sharp transitions occur—the system smoothly evolves to the ferromagnetic phase through a series of crossovers as $H_p$ increases.

Thus, the non-stoquastic catalyst softens the para-ferro transition by inducing an intermediate transition at smaller $s$, involving two low-magnetization states. This avoids a sharp magnetization jump and makes the gap closing less severe. Because the paramagnetic state is a superposition over all computational basis states, simple $k$-local catalysts cannot directly connect it to the ferromagnetic state unless all orders of $k$ are included. By initiating an energetic competition early in the anneal, the non-stoquastic catalyst breaks the symmetry of the paramagnetic state, introducing a new transition (with a gap closing as a power law in $L$) to a different intermediate state and then enabling a smooth crossover to the ferromagnetic ground state. This mechanism is fundamentally different from the one discussed in the main text, since it addresses a different source of gap-closing compared to the problems discussed in main text, and thus explains the differing roles played by the catalyst's sign.
}

\section{Coupling data for different examples in the main text}
\label{app:appC}
Below we give the parameters for the different instances of MWIS problem on random graphs which have been analyzed in different circumstances in the main text. If no coupling is given between a set of vertices it means the coupling is set to $0$.
\begin{table}[h]
    \centering
    \begin{tabular}{|c|c|c|c|}
        \hline
       \textbf{\( J_{ij} \)} & \textbf{Strength} & \textbf{\( J_{ij} \)} & \textbf{Strength} \\
        \hline
        $J_{1,4}$ & 1.66122 & $J_{1,6}$ & 1.01834\\
        $J_{1,8}$ & 1.14459 & $J_{2,3}$ & 1.78942 \\
        $J_{2,4}$ & 1.10915 & $J_{2,6}$ & 1.8282 \\      
        $J_{2,7}$ & 1.76385 & $J_{3,4}$ & 1.57587 \\
        
        $J_{3,9}$ & 1.03825 & $J_{3,10}$ & 1.88831 \\
        
        $J_{4,7}$ & 1.27207 & $J_{4,9}$ & 1.02395 \\
       
        $J_{4,10}$ & 1.68937 & $J_{5,6}$ & 1.23293 \\
       
        $J_{5,8}$ & 1.32764 &  $J_{5,9}$ & 1.0961 \\
         $J_{6,10}$ & 1.09028 & $J_{7,8}$ & 1.19425 \\
        
        $J_{7,9}$ & 1.22829 & $J_{7,10}$ & 1.96842 \\

        $J_{8,10}$ & 1.30031 & & \\
        \hline
    \end{tabular}
    \caption{Coupling strengths $J_{ij}$ used for the example in Fig.~\ref{fig:XXXbetter} in the main text}
    \label{tab:my_label}
\end{table}

\begin{table}
\centering
\begin{tabular}{|c|c|c|c|}
\hline
\textbf{\( J_{ij} \)} & \textbf{Strength} & \textbf{\( J_{ij} \)} & \textbf{Strength} \\
\hline
\( J_{1,6} \)  & 1.43618  & \( J_{4,6} \)  & 1.13057 \\
\( J_{1,7} \)  & 1.94469  & \( J_{4,7} \)  & 1.71423 \\
\( J_{1,8} \)  & 1.94228  & \( J_{4,9} \)  & 1.6577  \\
\( J_{1,9} \)  & 1.78754  & \( J_{4,10} \) & 1.50474 \\
\( J_{2,3} \)  & 1.14242  & \( J_{5,6} \)  & 1.97884 \\
\( J_{2,4} \)  & 1.9505   & \( J_{5,8} \)  & 1.154   \\
\( J_{2,5} \)  & 1.02682  & \( J_{5,9} \)  & 1.10829 \\
\( J_{2,6} \)  & 1.80606  & \( J_{5,10} \) & 1.35026 \\
\( J_{2,7} \)  & 1.44841  & \( J_{7,9} \)  & 1.35643 \\
\( J_{2,8} \)  & 1.09757  & \( J_{3,4} \)  & 1.73191 \\
\( J_{2,9} \)  & 1.51751  & \( J_{3,6} \)  & 1.7245  \\
\( J_{2,10} \) & 1.97066  & \( J_{3,10} \) & 1.0843  \\
\hline
\end{tabular}
\caption{Coupling strengths \( J_{ij} \) between connected nodes, for the red circle enclosed point in Fig.~\ref{fig:erdrenyicat}. }
\label{tab:redcoup}
\end{table}

\begin{table}
\centering
\begin{tabular}{|c|c|c|c|}
\hline
\textbf{\( w_i \)} & \textbf{Value} & \textbf{\( w_i \)} & \textbf{Value} \\
\hline
\( w_1 \) & 0.968722  & \( w_6 \) & 0.559234 \\
\( w_2 \) & 0.00924594 & \( w_7 \) & 0.904322 \\
\( w_3 \) & 0.440185  & \( w_8 \) & 0.19012 \\
\( w_4 \) & 0.113431  & \( w_9 \) & 0.786817 \\
\( w_5 \) & 0.667931  & \( w_{10} \) & 0.42323 \\
\hline
\end{tabular}
\caption{On-site potential values \( w_i \) for the red circle enclosed point in Fig.~\ref{fig:erdrenyicat}. }
\label{tab:redpot}
\end{table}

\begin{table}
\centering
\begin{tabular}{|c|c|c|c|}
\hline
\textbf{\( J_{ij} \)} & \textbf{Strength} & \textbf{\( J_{ij} \)} & \textbf{Strength} \\
\hline
\( J_{1,2} \) & 1.49747 & \( J_{4,6} \) & 1.98316 \\
\( J_{1,4} \) & 1.92165 & \( J_{4,7} \) & 1.19158 \\
\( J_{1,5} \) & 1.80992 & \( J_{4,8} \) & 1.7254 \\
\( J_{1,7} \) & 1.01589 & \( J_{5,6} \) & 1.85658 \\
\( J_{1,8} \) & 1.92879 & \( J_{5,8} \) & 1.87132 \\
\( J_{2,7} \) & 1.17451 & \( J_{5,10} \) & 1.03633 \\
\( J_{2,8} \) & 1.88529 & \( J_{6,8} \) & 1.64893 \\
\( J_{2,9} \) & 1.38427 & \( J_{7,8} \) & 1.55501 \\
\( J_{2,10} \) & 1.43671 & \( J_{7,9} \) & 1.73864 \\
\( J_{3,4} \) & 1.08609 & \( J_{9,10} \) & 1.58989 \\
\( J_{3,6} \) & 1.29071 & \( J_{3,10} \) & 1.34098  \\
\( J_{3,9} \) & 1.79068 &  &  \\
\hline
\end{tabular}
\caption{Coupling strengths \( J_{ij} \) between connected nodes, for the point circled in green in Fig.~\ref{fig:erdrenyicat}.}
\label{tab:greencoup}
\end{table}

\begin{table}
\centering
\begin{tabular}{|c|c|c|c|}
\hline
\textbf{\( w_i \)} & \textbf{Value} & \textbf{\( w_i \)} & \textbf{Value} \\
\hline
\( w_1 \) & 0.628552 & \( w_6 \) & 0.341512 \\
\( w_2 \) & 0.776818 & \( w_7 \) & 0.445503 \\
\( w_3 \) & 0.0318716 & \( w_8 \) & 0.354165 \\
\( w_4 \) & 0.63906 & \( w_9 \) & 0.463555 \\
\( w_5 \) & 0.0192283 & \( w_{10} \) & 0.501439 \\
\hline
\end{tabular}
\caption{On-site potentials \( w_i \)  for the point circled in green in Fig.~\ref{fig:erdrenyicat}.}
\label{tab:greenpot}
\end{table}

\rgg{\section{Improvement in Fidelity with continuous time anneal}
\label{app:appfinal}
\begin{figure}
    \centering
    \includegraphics[width=0.45\linewidth]{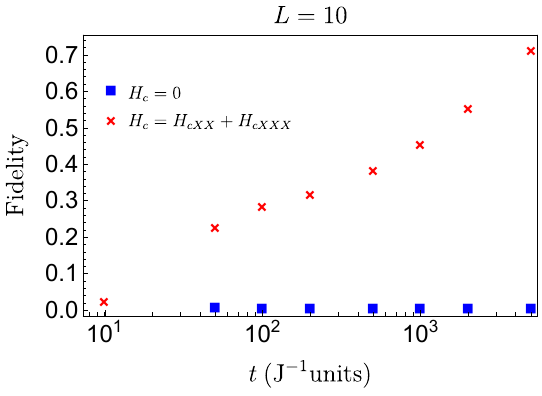}
        \includegraphics[width=0.45\linewidth]{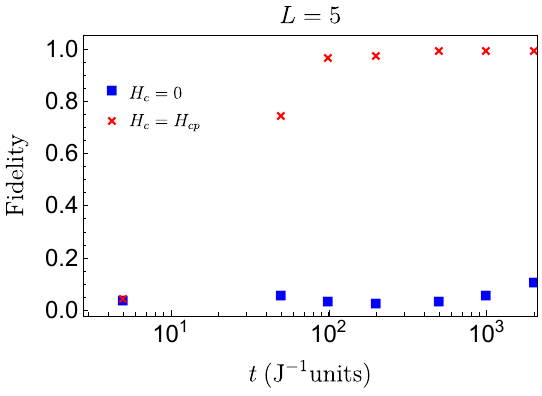}
    \caption{Left: Fidelity of annealed state with the true ground state for time $t$ in units of $J^{-1}$ for the example problem in Fig.~\ref{fig:XXXbetter}. Right: The same for the toy blicique model of $L=5$ with parameters of Fig.~\ref{fig:gap}.}
    \label{fig:continuouscatalyst}
\end{figure}
In Fig.~\ref{fig:continuouscatalyst}, we demonstrate that the use of $n$-local catalysts leads to an order-of-magnitude improvement 
in the time required to reach near-unity fidelity even in the continuous-time (adiabatic) regime. 
This result complements the discrete-time analysis presented in the main text and shows that 
the advantage conferred by the catalysts is not limited to the gate-based framework.

Importantly, our approach does not involve any optimization within a fixed catalyst configuration. 
For a given $n$-local catalyst, the annealing protocol is executed once, 
and the corresponding performance is directly evaluated. 
Hence, the additional overhead associated with exploring higher $n$ values 
amounts only to a linear correction in the catalyst locality, 
as opposed to an exponential search over configurations. 
This establishes that the overall time to identify an effective catalyst remains efficient, 
while still achieving substantial reductions in the total annealing time.}
\end{document}